\documentclass{emulateapj}

\begin{document}

\title{Population Synthesis of Black Hole X-Ray Binaries}

\author{
Yong Shao$^{1,2}$ and Xiang-Dong Li$^{1,2}$}

\affil{$^{1}$Department of Astronomy, Nanjing University, Nanjing 210046, People's Republic of China; shaoyong@nju.edu.cn}

\affil{$^{2}$Key laboratory of Modern Astronomy and Astrophysics (Nanjing University), Ministry of
Education, Nanjing 210046, People's Republic of China; lixd@nju.edu.cn}

\begin{abstract}

We present a systematic study of the X-ray binaries (XRBs) containing a black hole (BH) 
and a non-degenerate companion,
in which mass transfer takes place via either capturing the companion's wind or 
Roche lobe overflow (RLO). As shown in our previous work that focusing on 
the formation and evolution of  detached BH binaries, 
our assumed models relevant to BH's progenitors predicted significantly 
different binary properties  \citep{sl19}. 
In this paper, we further follow the evolutionary paths of the BH systems that appearing as 
XRBs. 
By use of both binary population synthesis and detailed binary 
evolution calculations, we can obtain the potential population of BH XRBs. 
Distributions at the current epoch of various binary parameters have been computed. 
The observed sample of wind-fed XRBs can be well reproduced under assumption of all our models.
The wind-fed XRBs are expected to be so rare ($ \lesssim 100 $) that only a couple of such systems
have been detected. 
Comparison of known RLO XRBs with the calculated distributions of various binary parameters
indicates that only the models assuming relatively small masses for
BH progenitors can roughly match the observations. 
Accordingly we estimate that there are hundreds of RLO XRBs in the Milky Way, 
of which the majority are low-mass XRBs. The RLO systems may become ultraluminous X-ray sources (ULXs) if the BH accretes 
at a very high rate, we expect that about a dozen ULXs with a BH accretor may exist in a Milky Way$-$like galaxy.

\end{abstract}


\section{Introduction}

To date, there are about two dozen of binary systems containing black holes (BHs) for which 
their dynamical masses have been confirmed in the Milky Way. The majority of them are identified in X-ray binaries
(XRBs), since the X-ray radiation due to the BH accretion was detected \citep{rm06,cj14}.  Prior to
the XRB phase, the systems ought to be detached binaries hosting a quiescent BH around its 
non-degenerate companion. Recently, a few detached systems are proposed to host a BH since its dynamical
mass was measured according to the optical observations of the visible companion \citep[e.g.,][]{gd18,km18,tk19,liu19}.

\begin{table*}
\begin{center}
\caption{Basic parameters of Galactic (candidate) BH XRBs with known orbital periods. 
\label{tbl-1}}
\begin{tabular}{lcccccccclllll}
\\
\hline
Source  & $ M_{\rm BH} $ ($ M_{\odot} $) & $ M_{\rm c} $  ($ M_{\odot} $) & 
$ P_{\rm orb} $ (days) & Spectral type & $ T_{\rm eff} $ & $ a_{\ast} $ & $ \dot{M}_{\rm tr} $ 
($ M_{\odot}\,\rm yr^{-1} $) \\
\hline
GRS 1915+105 (1) & $ 12.4 \pm 2.0 $     &  $0.58\pm 0.33$    & 33.85 & K0III$ -$K3III & $ 4100-5433 ^{c}$ & $0.95 \pm 0.05$ & $   <  1.5 \times10^{-7} $\\
GS 2023+338 (2) & $ 9.0 \pm 0.6 $     &  $0.54\pm 0.05$    & 6.47 & K0III$ -$K3III & $ 4100-5433 ^{c}$ & $-$ & $   9.3 \times10^{-10} $\\
V4641 Sgr (3) & $ 6.4 \pm 0.6 $     &  $2.9\pm 0.4$    & 2.82 & B9III & $ 10500\pm 200 $ & $-$ & $ - $\\
GRO J1655-40 (4) & $ 5.31 \pm 0.07 $ & $1.75\pm 0.25$  & 2.62 & F3-G0IV & $ 5706-6466^{c} $& $0.7\pm 0.1$ & $ 8.7 \times10^{-10}$\\
GS 1354-64 (5) & $ 7.6 \pm 0.7 $     &  $\sim1.03$    & 2.54 & G5IV & $ 4985-6097^{c} $& $-$ & $ > 1.6 \times10^{-9}$\\
GX 339-4 (6) & $ > 7$     &  $0.3-1.1$   & 1.75 & K1V & $4533 - 4837^{c} $ & $0.25\pm 0.15^{a}$ & $  1.8 \times10^{-8}$\\
XTE J1550-564 (7) & $ 9.1 \pm 0.6$   &  $0.30\pm 0.07$ & 1.54 & K3III/V & $ 4700\pm 250 $ & $0.34\pm 0.2$ & $  1.6 \times10^{-9}$\\
4U 1543-47 (8) & $ 9.4 \pm 2.0$   &  $2.7\pm 1.0 $ & 1.12 & A2V & $ 9000\pm 500 $ & $0.8\pm 0.1$ & $  1.1 \times10^{-9}$\\
MAXI J1820+070 (9) & $5.7-8.3$ & $0.3-0.8$ & 0.69 & K3V-K5V & $ 4065-5214^{c} $ & $-$ & $-$ \\
H1705-250 (10) & $ 6.4 \pm 1.5$   &  $0.25\pm 0.17 $ & 0.52 & K3V-M0V & $ 3540-5214 ^{c}$ & $-$ & $  < 2.8 \times10^{-10}$\\
GRS 1124-68 (11) & $ 6.95 \pm 0.6$  & $0.9\pm 0.3 $ & 0.43 & K3V-K5V & $ 4065-5214^{c} $ & $0.25\pm 0.15 ^{a}$ &$ < 3.3 \times10^{-10}$\\
XTE J1859+226 (12) & $ 7.7 \pm 1.3$  & $\sim 0.7$ & 0.38 & K0V-K5V & $4400 -5240^{c} $& $0.25\pm 0.15 ^{a}$ &$ - $\\ 
GS 2000+251 (13) & $ \sim 6.55$  & $0.16-0.47$ & 0.34 & K3V-K6V & $ 3915-5214^{c} $ & $0.05\pm 0.05 ^{a}$ &$ < 6.5 \times10^{-11}$\\ 
A0620-00 (14)  & $ 6.61 \pm 0.25$  & $0.4\pm 0.05$ & 0.32 & K5V-K7V & $ 3800-4910^{c} $ & $0.12\pm 0.19 $ &$ 5.4 \times10^{-11}$\\
GRS 1009-45 (15) & $ \sim 8.5$  & $0.54\pm 0.10$ & 0.29 & K7V-M0V& $ 3540-4640^{c} $ & $-$ &$ 2.7 \times10^{-10}$\\
GRO J0422+32 (16) & $ \sim 10.4$  & $0.95\pm 0.25$ & 0.21 & M1V-M4V & $ 2905-4378^{c} $ & $-$ &$ < 2.7 \times10^{-11}$\\
XTE J1118+480 (17) & $ 7.6\pm 0.7$  & $0.18\pm 0.07$ & 0.17 & K7V-M1V & $ 3405-4640^{c} $& $-$ &$  2.1 \times10^{-10}$\\
MAXI J1659−352 (18) & $ 3.6-8.0$  & $\sim 0.15-0.2$ & 0.1 & M2V-M5V & $ 3200-3600^{c} $& $-$ &$  1.4 \times10^{-10}$\\
\\
Cyg X-1 (19)  &     $ 14.8 \pm 1.0 $  & $ 19.2\pm 1.9 $   & 5.6 & O9.7Iab & $ - $& $> 0.983  $ & $  1.4 \times 10^{-8}$\\
Cyg X-3 (20)  &   $ 2.0-4.5 $  &  $7.5-14.2 $   & 0.2 & WN4$-$WN8 & $ - $ & $ - $ & $  - $\\
AS 386$ ^{b}$  (21)   & $ >7  $     & $ 6-8 $   & 131.3 & Be  & $-$ & $-$ & $ -$\\
MWC 656$ ^{b}$ (22)  & $ 3.8-6.9 $     & $ 10-16 $   & 60.4 & B1.5-B2IIIe & $- $ & $-$ & $ -$\\
\hline
\end{tabular}
\end{center}
Column 1: the source name. Column 2: the BH mass. Column 3: the companion mass. Column 4: the orbital period.
Column 5: the companion's spectral type. Column 6: the companion's surface effective temperature. 
Column 7: the BH's spin parameter. Column 8: the mean mass-transfer rate, 
which is obtained according to the estimation of \citet{cfd12}.\\
Notes. $ ^{a} $ The $ a_{\ast} $  is estimated from the maximum jet power of the source \citep{sm13}. 
$ ^{b} $ AS 386 and MWC 656 do not belong to XRBs since they are undetected in the X-ray band, the binary
parameters listed here will be used to explore the properties of the BH$-$Be binaries in Section 2.1. 
$ ^{c} $ The $ T_{\rm eff} $ is inferred from the given spectral type. \\
References. (1) \citet{gcm01}, \citet{hg04},
\citet{msn06}. (2) \citet{cc94}, \citet{kfr10}, \citet{hbr09}. (3) \citet{okv01}, \citet{saa06}, \citet{mbb14}.
(4) \citet{mbs14}, \citet{gh08}, \citet{smr06}. (5) \citet{coz09}. (6) \citet{mcm08}, \citet{hsc03}. (7) \citet{os11},
\citet{sr11}. (8) \citet{oj98}, \citet{smr06}.  (9) \citet{tc19}, \citet{tc20}.
(10) \citet{ro96}, \citet{hs97}. (11) \citet{sn97}, \citet{gh01}.
(12) \citet{zs02}, \citet{cs11}. (13) \citet{ir04}, \citet{hh96}, \citet{ci90}. 
(14) \citet{jp09},  \citet{cb10}, \citet{ns08}, \citet{gm10}. (15) \citet{fl99}, \citet{mo11}. 
(16) \citet{hc99}, \citet{rc07}. (17) \citet{gh12}, \citet{kfr13}, \citet{cv09}. (18) \citet{ya12}, \citet{kk13}.
(19) \citet{om11}, \citet{gmr14}. 
(20) \citet{zmb13}, \citet{vk92}. (21) \citet{km18}. (22) \citet{cnr14}. 
\end{table*}

The mass transfer proceeds through either the capture of the companion's wind 
or Roche lobe overflow (RLO), XRBs can be accordingly divided into 
the wind-fed systems and the RLO ones. Several wind-fed XRBs 
have been confirmed to contain a BH accretor and an OB star companion 
such as Cyg X-1 \citep[e.g.,][]{om11} and LMC X-1 \citep[e.g.,][]{os09}. 
Also, in Cyg X-3 the compact star is very likely a BH \citep{zmb13} that
is  being fed by the wind from a Wolf-Rayet (WR) star \citep{vk92}. While  
the wind-fed systems usually belong to high-mass ($ \gtrsim 10 M_\odot $) XRBs (HMXBs),
observations show that most of BH XRBs are the RLO systems with companion masses less than $ \sim 2 M_\odot $
\citep{rm06,cj14}.  These binary systems are correspondingly classified as low-mass XRBs (LMXBs).  
Table 1 provides the observed values of key parameters of known BH XRBs in the Milky Way.
Among the observed XRB sample, it seems to be lack of intermediate-mass ( $ \sim 3-10 M_\odot $)
XRBs (IMXBs). On the one hand, as a wind-fed system, the stellar wind from an intermediate-mass 
companion is not strong enough, thus such a source is too dim to be detected in the X-ray band. 
On the other hand, as an RLO system, the mass transfer will proceed on a very short timescale, 
leading to the rarity of the IMXB systems. 
SS433 might be in such a stage with super-Eddington mass transfer, although its nature is still of debate \citep[see][for a review]{f04}. 
Meanwhile, in external galaxies, there are hundreds of ultraluminous X-ray sources (ULXs) observed, whose X-ray luminosities 
exceed $ 10^{39}\,\rm erg\,s^{-1} $ \citep{kfr17}. It is thought that such high X-ray luminosities are created 
due to mass accretion onto BHs in the RLO XRBs \citep{prh03,rpp05}. Previous population synthesis 
calculations showed that the ULX systems with a BH accretor are dominated by HMXBs and IMXBs,
whose ages are always younger than 100 Myr  \citep[e.g.,][]{mr08,ws17}. 
Besides the BH ULXs
in a young environment, a part of the ULX systems are expected to be LMXBs in old populations \citep{k02}. 

The general picture for the formation and evolution of BH XRBs has been built for decades \citep{vh74}.
BH XRBs are thought to be the evolutionary products of massive primordial binaries.
However, there still remain
many major uncertainties in the processes of massive binary evolution, in particular the physics of 
the BH formation and the role of the mass exchange \citep[e.g.,][]{pjh92,l12}. Many previous investigations
have explored the formation of the BH HMXBs containing an OB star 
\citep[e.g.,][]{bz09} or a helium star \citep{lvk12}, and demonstrated that the possible properties of the
XRB systems are significantly subject to those uncertainties. The HMXB stage plays a vital role in connecting
the evolution from massive primordial binaries to double compact stars. One expects that
BH HMXBs may evolve to be double compact stars if the binary systems are not disrupted
or merged during the evolution, and a part of them having close orbits
may finally coalesce into gravitational wave sources \citep[e.g.,][]{bh16,kt18,mg18}.

The formation of BH LMXBs is still a controversial topic. 
According to conventional wisdom, the BH LMXBs are believed to evolve from 
the primordial binaries with extreme mass ratios \citep[see][for a review]{l15}. 
The standard scenario for the BH LMXB formation involves a common envelope (CE) phase \citep{p76} 
during the primordial binary evolution. 
When the secondary star and the core of the primary star avoid merging, the core may evolve to
collapse and finally lead to the formation of a BH in a close binary. One obvious difficulty in this scenario is that
the orbital energy of the binary system may be insufficient to eject the massive envelope of the primary
star \citep[e.g.,][]{p97,k99,prh03}. 
This has led to several exotic scenarios for the BH LMXB formation, e.g., a triple star scenario \citep{ev86},
the formation of the low-mass companion from a disrupted envelope of a massive star \citep{p95}, 
the companion being a
pre-main sequence star \citep{i06}, or the ejection of the primary's envelope with nuclear rather than 
orbital energy \citep{p10}. In addition, it was suggested that the current LMXBs may descend from the BH binaries 
with an initially intermediate-mass companion \citep{j06,cl06,l08}.

The birthrates and properties of BH XRBs are strongly influenced by the condition of BH formation.
Stellar evolution predicts that BHs are the final products of massive stars with masses $ \gtrsim 20-25M_\odot $
\citep{ww95,f12}. However, there are observational indications that stars of mass $\gtrsim 17-25 M_\odot $ 
have not been observed as the progenitors of type IIP supernovae \citep{s09,s15}, 
\citet{k14} suggested that these stars may die in failed supernovae creating BHs. The
formation problem of BH LMXBs may be solved in the standard scenario if the lower limit for the masses of
the BH progenitors is down to $ \sim 17 M_\odot $ \citep{wjl16}, which help the binary survive during the 
CE evolution. Recent numerical simulations \citep[e.g.,][]{oo11,uj12,pt15,ej16,se16} indicate
that the outcome of neutrino-driven explosions is greatly controlled by the core structure of massive stars, 
and there is no clear threshold of the progenitor masses to determine whether the stars implode to be BHs
or explode to be neutron stars (NSs). Stars with masses even as low as $ \sim 15M_\odot $ can still have the probability  
to eventually implode to be BHs \citep{rs18}. To explore the possible influence of different BH formation mechanisms, 
\citet[][hereafter Paper I]{sl19} proposed four relevant models in their investigation in the
population of detached BH binaries with normal-star companions\footnote{A normal star 
specifically refers to a star residing at the main-sequence or (super)giant stage.} 
in the Milky Way. 
These BH binaries, although quiescent in X-ray, can be detected and dynamically identified by observations 
of the optical companions \citep{gd18,km18,tk19,liu19}.

In paper I, we use the binary population synthesis (BPS) code BSE  \citep{h02} to obtain 
the distributions of various parameters for incipient BH binaries\footnote{An incipient BH binary is a 
binary system containing a newly born BH and an unevolved main-sequence star.}, 
including the component masses, the evolutionary state of the companion stars and the orbital parameters 
of the binary systems. When the winds from the companion star become intense enough, or when the companion 
star evolves to overflow its RL, efficient mass accretion onto the BH commences and the binary appears as an XRB.
In this work, we attempt to explore the properties of the XRB population including both the 
wind-fed binaries and the RLO ones based on the results in Paper I. 
Since the BSE code can model the mass transfer via RLO in a rather crude way, we employ the stellar 
evolution code \textit{MESA} \citep{p11,p13,p15} to simulate the mass transfer processes in RLO systems,
by tracking the evolutionary paths of several thousands of the BH binaries with a grid of initial parameters. 
After recording the evolutionary sequences of each BH binary during the mass-transfer phases, we can combine
them with the birthrate distribution of the incipient BH binaries to
synthesize the potential population of the RLO XRBs. With the \textit{MESA} code to simulate the
evolution of the lobe-filling binaries, this is a major step forward over previous
population synthesis studies of the RLO systems. 

The structure of this paper is organized 
as follows.  In Section 2 we use the BPS method to show the properties of the wind-fed XRBs,
and then employ detailed binary evolution calculations to obtain the outcomes of the RLO XRBs in Section 3.
We briefly discuss some possible uncertainties that can affect our results in Section 4.
Finally we make a conclusion in Section 5.

\begin{table*}
\begin{center}
\caption{Predicted numbers of various types of BH XRBs in the Milky Way.
\label{tbl-2}}
\begin{tabular}{cccccccclllll}
\\
\hline
Models     & $ M_{\rm BH} $ & $V_{\rm k}$ & $ N_{\rm BH-Be}$ & $  N_{\rm BH-OB}^{\rm wind-fed} $ & $ N_{\rm BH-He}^{\rm wind-fed} $ & $ N_{\rm XRB}^{\rm RLO} $ &  $ N_{\rm ULX}^{\rm RLO} $ \\
\hline
A   &  $ M_{\rm proto} +  M_{\rm fb} $      & $ \propto1-f_{\rm fb} $   & 378& 34 (113)& 81 (200) & 53 & 12\\
B     & $ M_{\rm rem} $                & $ \propto 3.0/M_{\rm BH} $  & 392 & 19 (156) & 56 (342)& 586 &  16 \\
C     &$ M_{\rm rem} $       &  $ \sigma_{\rm k} = 150\,\rm km\,s^{-1} $    & 332 & 13 (153) & 34 (364) & 622&  14 \\
D     &  $ M_{\rm rem} $     & $ \sigma_{\rm k} = 50\,\rm km\,s^{-1} $ &  1044  & 30 (444)& 109 (542)& 822  & 17 \\
\hline
\end{tabular}
\end{center}
Columns 1-3: the physical inputs of different models. Column 4: for all detached BH$-$Be binaries, in 
spite of the magnitudes of the X-ray luminosity due to BH accretion. Columns 5-6: for the 
wind-fed BH$-$OB and BH$-$He XRBs with X-ray luminosities of $\geqslant 10^{35}\,\rm erg\,s^{-1}$.  
For comparison, the corresponding numbers of all detached binaries are given in the parentheses.  
Column 7: for the RLO XRBs. Column 8: for the RLO ULXs. \\
Notes. In this study, we have separated the BH systems with an OB star into BH$-$Be binaries and BH$-$OB ones,  
depending on whether or not the rotational velocities of the OB star reach $ 80 \%$ of the Keplerian limits. 
\end{table*}

\section{Populations of Wind-fed XRBs}

Observations of Galactic BH XRBs show that the wind-fed systems are usually HMXBs. 
According to different stellar types of the companion star, the wind-fed XRBs are
divided into the binary systems containing an OB star and the ones containing 
a helium star (or a WR star).

Before describing the detailed properties of the calculated XRB populations, 
we review some basic assumptions and initial parameters used in Paper I.
If the mass transfer via RLO is dynamically unstable, then a binary will enter a CE phase.
We employ the standard energy conservation equation \citep{w84} to deal with the outcome of the CE phase.
It is assumed that the orbital energy of the embedded binary is used to eject the envelope. 
We adopt the results of \citet{xl10a,xl10b} to calculate the binding energy of the envelope and take the CE
ejection efficiency to be unity.

For BH formation, we adopted four different models regarding
the BH progenitors and the kick velocity distribution of the newborn BHs. 
Model A uses the prescription for the BH
formation via the rapid supernova mechanism \citep{f12}, in which single stars with masses of 
$ \gtrsim 20 M_\odot $ finally evolve to be BHs. The BH mass 
is contributed by the masses of both the proto-compact object and the fallback material.
For the distribution of BH's natal kick velocities, we use the velocity distribution of the
Galactic pulsars reduced by a factor of $ (1-f_{\rm fb} )$, where $ f_{\rm fb} $ is the fraction 
of the fallback material in the total ejected matter. The kick velocities of the Galactic pulsars are assumed to obey a Maxwellian 
distribution with a dispersion of $ 265\,\rm km\,s^{-1} $ \citep{h05}. 
In Models B$-$D, the progenitor 
masses of the BHs are allowed to be as low as $ \sim 15M_\odot $, and the BH masses 
are directly obtained from the remnant masses of the helium core prior to supernova 
explosions according to \citet{rs18}. There is a caveat that the lower-limit mass of $ \sim 15M_\odot $
for BH progenitors is worked out according to numerical simulations of single-star evolution \citep[e.g.,][]{se16}, 
while the final fate of stars in binary systems can be dramatically different due to RLO mass transfer. Massive (up to
$ \gtrsim 60 M_\odot $) stars that lose their envelopes before helium core burning end up with much lighter pre-supernova cores 
than those do not, and probably evolve to
be NSs rather than BHs \citep{b01}. Therefore, the models of single stars are only relevant for Case C binaries in which
mass transfer begins after helium core burning. From the BPS outcome, the primordial binaries 
with primary masses of $ \sim 15-20M_\odot $ always evolve to be the incipient BH binaries with a low-mass
companion, and almost all of them have experienced Case C mass transfer and then a CE phase. 
So adopting the single-star models to form BHs should be reasonable treatments in our cases.
Since the BH formation does not involve
the fallback process in Models B$-$D, we adopt three 
approaches to deal with the BH's natal kicks. In Model B we use the pulsar's kick velocity reduced 
by a factor of $ (3M_\odot/M_{\rm BH} )$. In Model C and D we adopt the kick velocities 
obeying a Maxwellian distribution with a dispersion of 150 and $ 50\,\rm km\,s^{-1}$, respectively.

For the primordial binaries, the distributions of all initial parameters are set as same as those in Paper I. 
We assume that the star formation rate of the Milky Way has a constant value of 
$ 3 M_{\odot}\,\rm yr^{-1} $ over a period of 10 Gyr and all stars initially have the solar compositions ($ Z = 0.02 $).
In our calculations, we pick out all of detached BH binaries and record relevant parameters at each of the
evolutionary step. The corresponding number of a specific binary can be calculated by multiplying its birthrate
with the timestep. Stellar wind mass-loss
rates of \citet{vink01} are applied to hot OB stars. For helium stars (or WR stars), we decrease the mass-loss 
rates of \citet{ham95} by a factor of 2 \citep{kh06}.
We follow the method of \citet{bk08} to calculate the mass accretion rates of BHs in the wind-fed systems. 
For eccentric binaries we estimate a mean accretion rate over the orbital period.  
The release of
the gravitational potential of the accreted material is converted into the X-ray luminosity. 

\begin{figure*}[hbtp]
\centering
\includegraphics[width=0.8\textwidth]{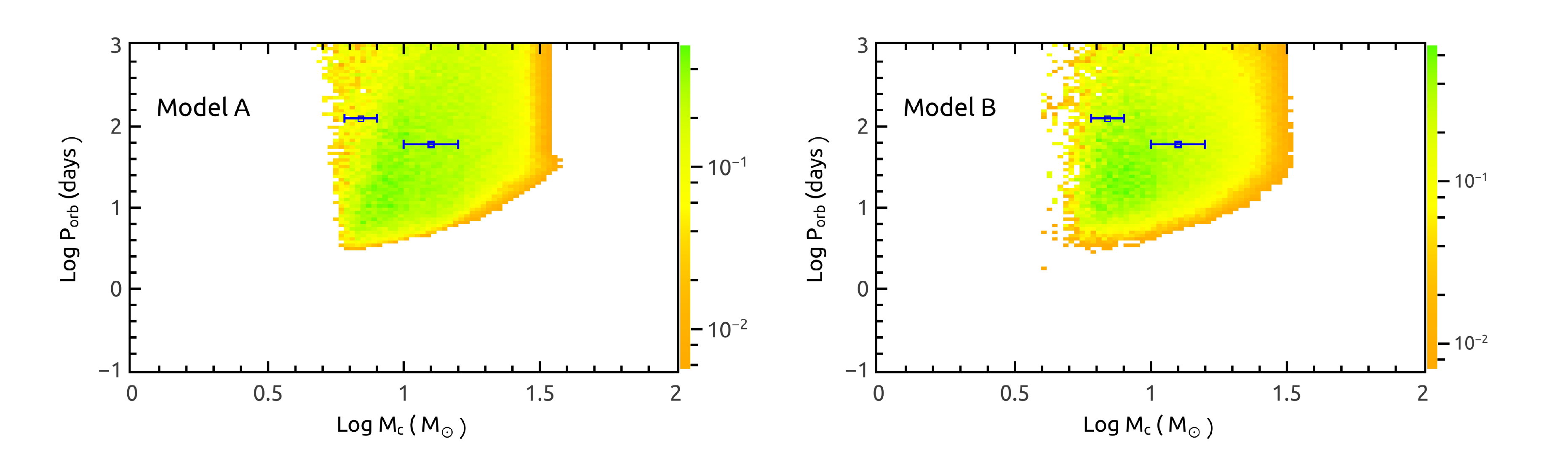}
\includegraphics[width=0.8\textwidth]{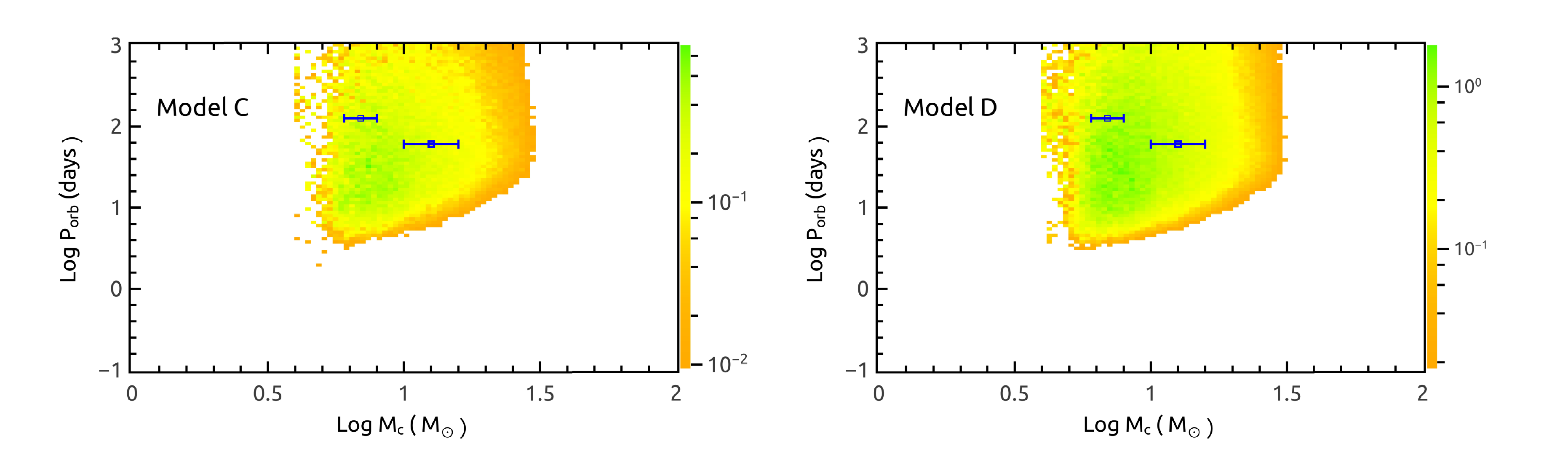}
\caption{Predicted number distributions of Galactic detached BH$-$Be binaries in the companion mass$-$orbital
period plane for Models A$-$D. The colors in each pixel are scaled 
according to the corresponding
numbers. The two square symbols in each panel mark the positions of the sources MWC 656 and AS 386. 
   \label{figure1}}
\end{figure*}

\begin{figure*}[hbtp]
\centering
\includegraphics[width=0.75\textwidth]{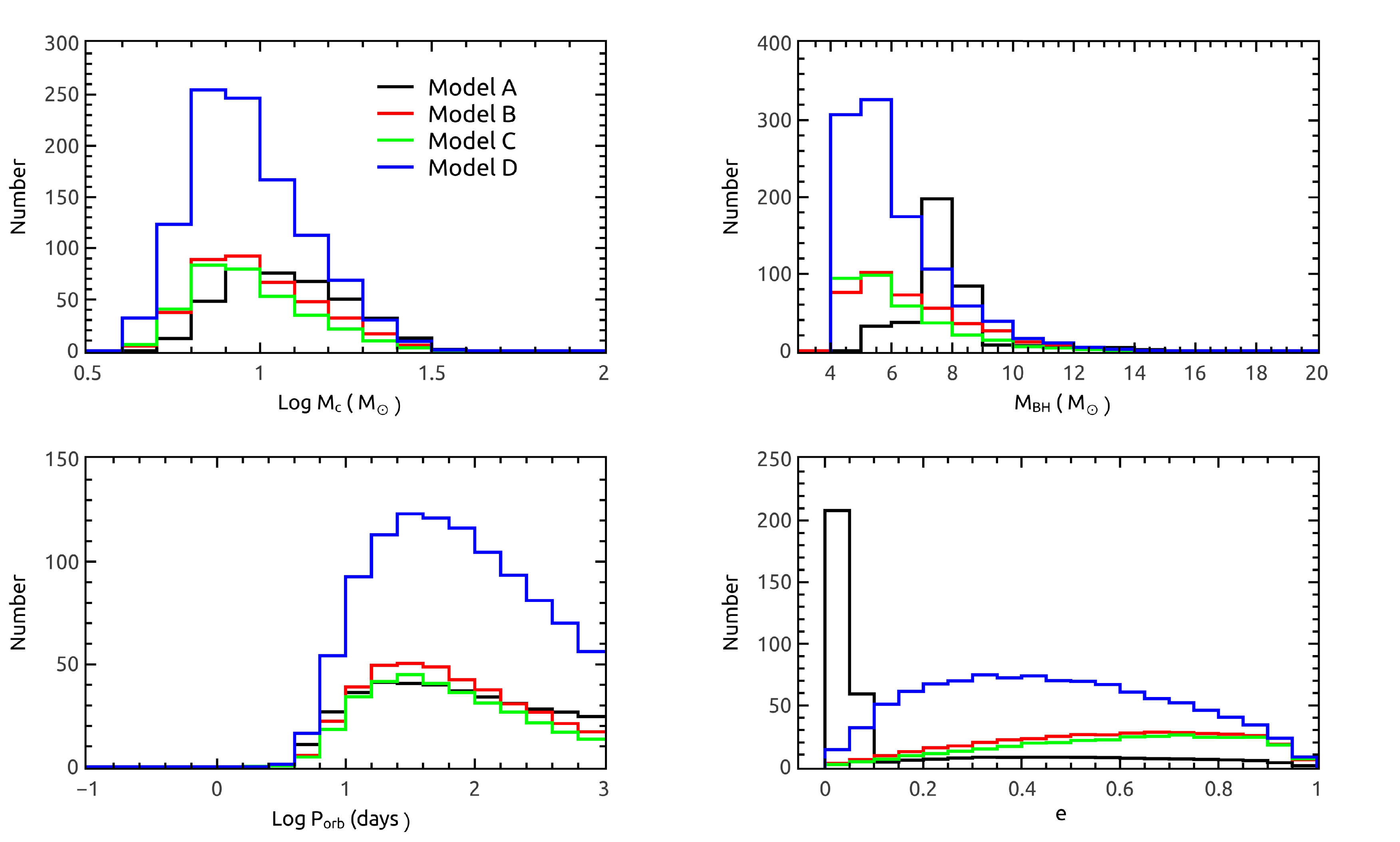}
\caption{Calculated number distributions of all detached BH$-$Be binaries in the Milky Way as a function of the companion
mass, the BH mass, the orbital period, and the orbital eccentricity. 
   \label{figure2}}
\end{figure*}

\begin{figure*}[hbtp]
\centering
\includegraphics[width=0.8\textwidth]{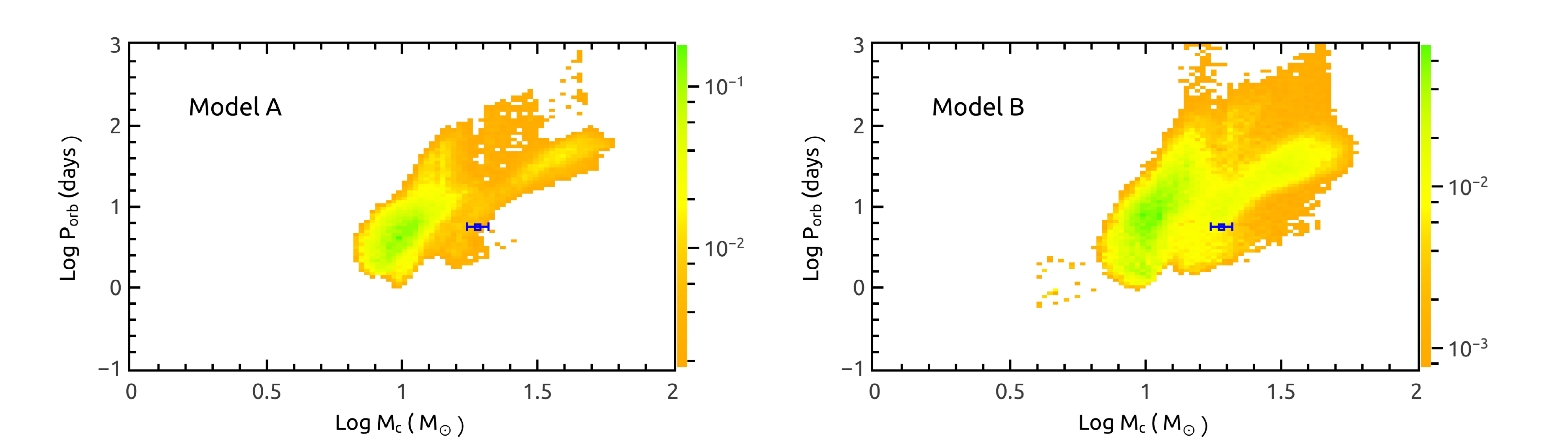}
\includegraphics[width=0.8\textwidth]{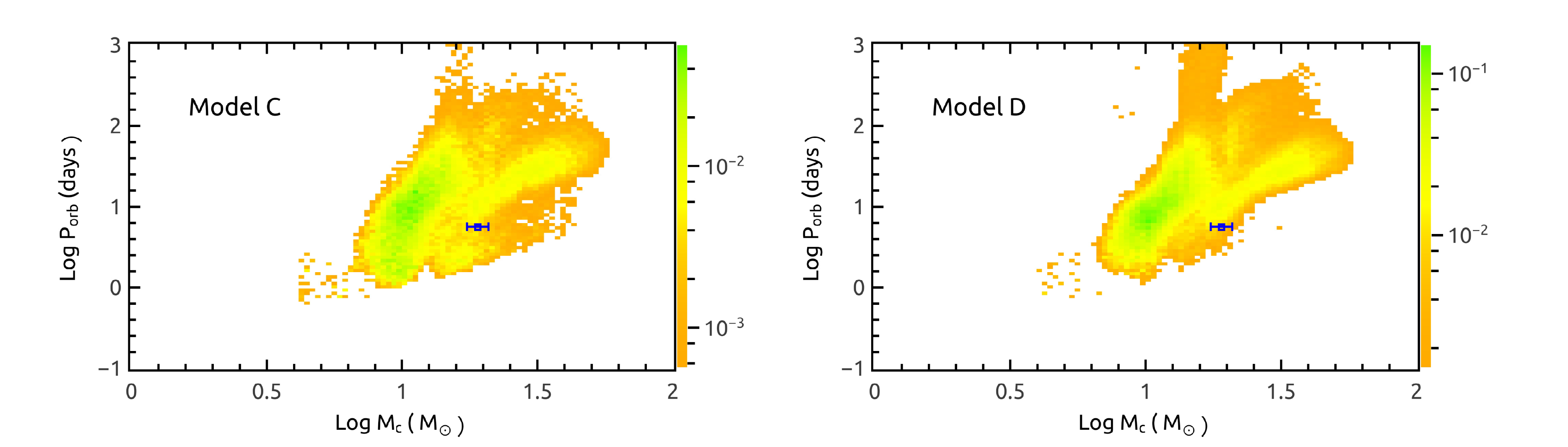}
\caption{Predicted number distributions of Galactic wind-fed BH$-$OB XRBs 
in the companion mass$-$orbital
period plane for Models A$-$D. The colors in each pixel are scaled 
according to the corresponding
numbers. The square symbol in each panel marks the position of the source Cyg X-1. 
   \label{figure3}}
\end{figure*}


Among the HMXBs with an NS accretor, most of them have been found to contain a Be star companion
\citep{lvv06,r11}. Be stars are thought to be rapidly rotating B-type stars. The rotational velocities 
of Be stars generally 
reach nearly their Keplerian limits, resulting in the formation of an excretion disk around them 
\citep[e.g.,][]{los91,msj13}. In an NS$-$Be XRB, 
the NS accretes the dense stellar wind from the Be star, therefore radiating X-rays. However, 
no BH$-$Be XRBs have been detected until now. MWC 656 and AS 386 are proposed to be the
BH$-$Be systems due to the optical (radial velocity) observations of the Be star \citep{cnr14,km18}. 
In this study, we separately discuss the wind-fed XRBs according to the 
rotational velocities of the companion star.
Using $ 80 \%$ of the Keplerian rotational velocity as a criterion, 
we separate BH binaries with an OB star into BH$-$Be systems and BH$- $OB ones, which 
respectively contain a rapidly and slowly rotating OB stars. In addition, we also take into account the 
BH binaries with a helium star (denoted as BH$-$He systems) when the hydrogen envelope of the companion was stripped off. 
In this work, we define the wind-fed systems as XRBs if the X-ray luminosities are higher than 
$ 10^{35}\,\rm erg\,s^{-1}$. The estimated numbers of  
wind-fed BH XRBs in all our assumed models are summarized in Table 2.  

\subsection{The Case of BH$-$Be Systems}

Since we cannot simulate the stellar winds from Be stars and then the X-ray luminosities due to BH accretion, 
here we consider all detached BH$-$Be systems.
Fig.~1 presents the predicted number distributions of Galactic BH$-$Be binaries in the companion mass
vs. orbital period plane. The four panels correspond to Models A$-$D. 
The two blue squares in each panel plot the positions of the sources MWC 656 and AS 386.
It can be seen that the covered region 
of the BH$-$Be binaries produced in all models are very similar. 
The masses of the companion stars vary in the range of $ \sim 5-30M_\odot $,
and the orbital periods are always larger than $ \sim 4 $ days. Note that only the binaries with orbital periods 
less than 1000 days are included in our study. The formation of Be stars in binary systems usually 
involves a stable mass-transfer stage rather than a CE phase during the progenitor binary evolution  
\citep[e.g.,][]{sl14,vnv20}. 
Our previous work on BH$-$Be binaries obtained slightly larger value ($ \sim 8M_\odot $) 
for the lower limit of the Be star masses \citep{sl14}. This difference of the companion masses
may originate from the different prescriptions of the stellar winds. Our present work allows lower masses for the BH's
progenitors since a weaker wind in the WR stage is adopted, thus leading to 
the formation of the Be star with smaller masses. The observed sources MWC 656 and AS 386
can be well reproduced in the parameter space of companion mass vs. orbital period. We estimate
that the total number of the BH$-$Be binaries in the Milky Way is $ \sim 300-1000 $ under assumption of all our  
models (see Table 2).

\begin{figure*}[hbtp]
\centering
\includegraphics[width=0.75\textwidth]{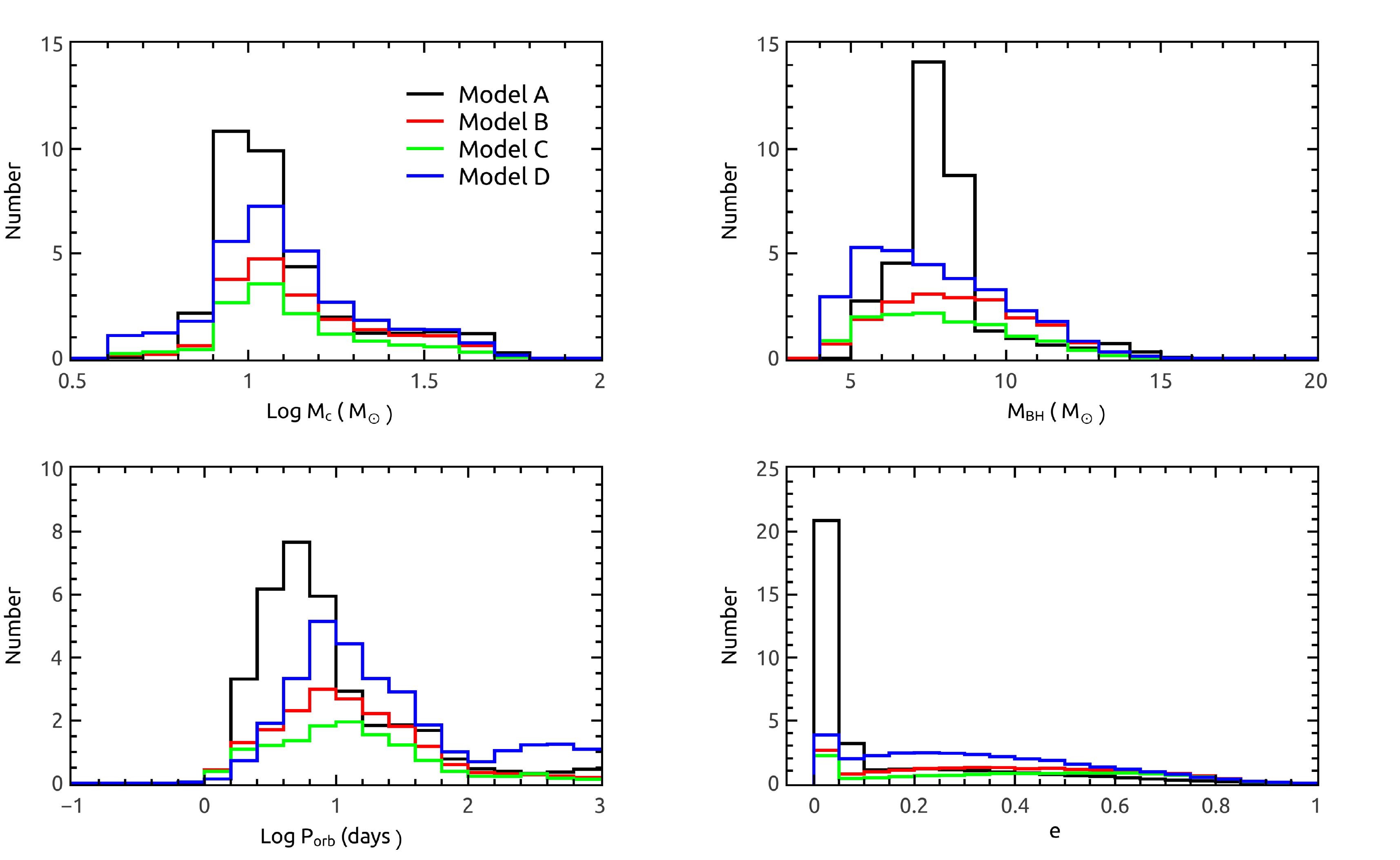}
\caption{Calculated number distributions of Galactic wind-fed BH$-$OB XRBs
 as a function of the companion
mass, the BH mass, the orbital period, and the orbital eccentricity. 
   \label{figure4}}
\end{figure*}

Fig.~2 shows the histogram distributions of the calculated number of Galactic BH$-$Be binaries as a function of 
the companion mass, the BH mass, the orbital period, and the orbital eccentricity in Models A$-$D. we can see that
the Be star masses and the orbital periods in Models A$ - $C have similar distributions, 
while Model D predicts three times more BH$ - $Be systems due to small BH's natal kicks. All our models indicate
that the mass distributions of the Be stars have a broad peak near $ \sim 6-10M_\odot $ and the orbital period distributions 
have a peak at $ \sim 30 $ days. 
The obvious differences between Models A and B$ - $D are the distributions of the BH masses and the orbital eccentricities.  
Model A predicts that the BH masses distribute in the range of $ \sim 5-15M_{\odot} $ with a peak at 
$ \sim 7-8M_{\odot}$, and the binary systems tend to have nearly circular orbits. This is because the majority
of the BHs are created through direct collapse without natal kicks. In Models B$ - $D,  the BH mass distributions
have a peak at $ \sim 5-6M_\odot $ and a tail up to $ \sim 15M_\odot $, more binaries tend to harbour lighter
BHs due to the IMF. Meanwhile, the BH$-$Be systems tend to have large eccentricities in Models B$ - $C
and modest eccentricities in Model D. 

\begin{figure*}[hbtp]
\centering
\includegraphics[width=0.8\textwidth]{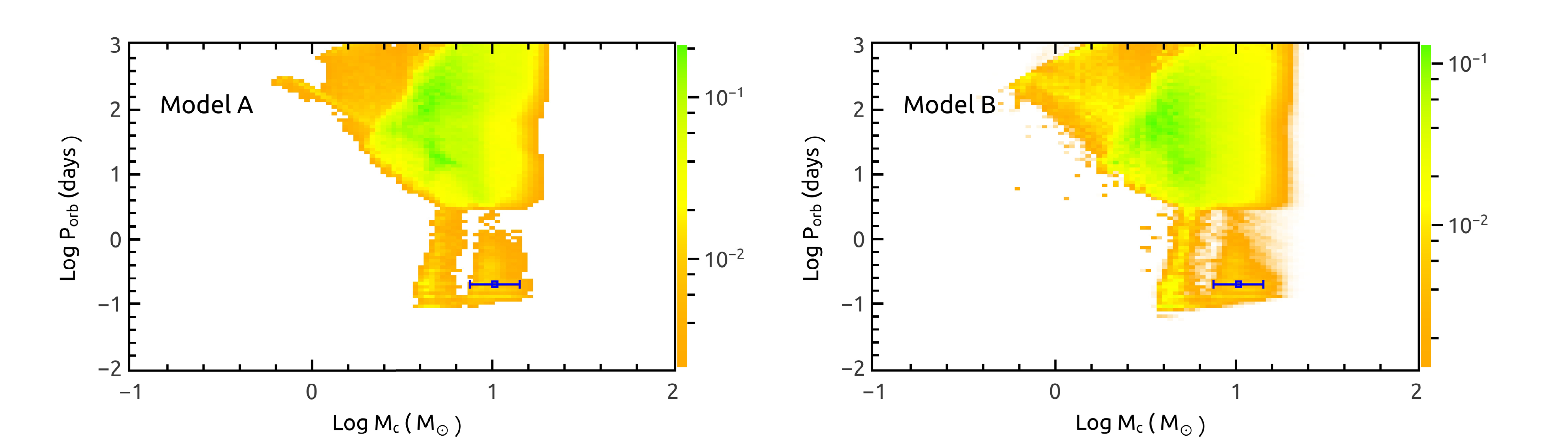}
\includegraphics[width=0.8\textwidth]{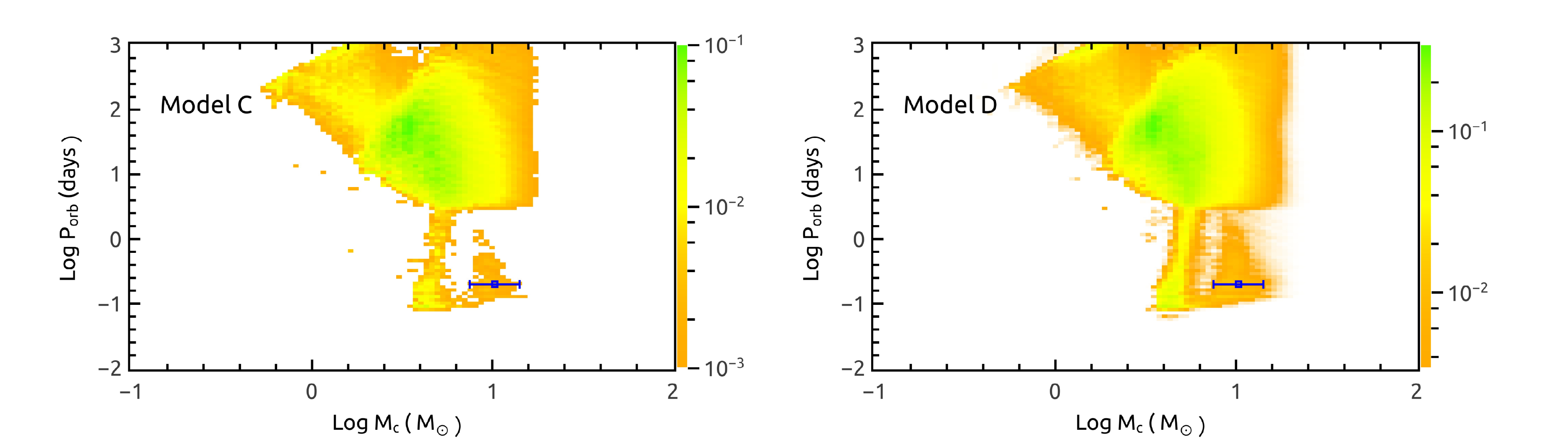}
\caption{Predicted number distributions of Galactic wind-fed BH$-$He XRBs
in the companion mass$-$orbital
period plane for Models A$ - $D. The colors in each pixel are scaled 
according to the corresponding
numbers. The square symbol in each panel marks the position of the source Cyg X-3. 
   \label{figure5}}
\end{figure*}

\begin{figure*}[hbtp]
\centering
\includegraphics[width=0.75\textwidth]{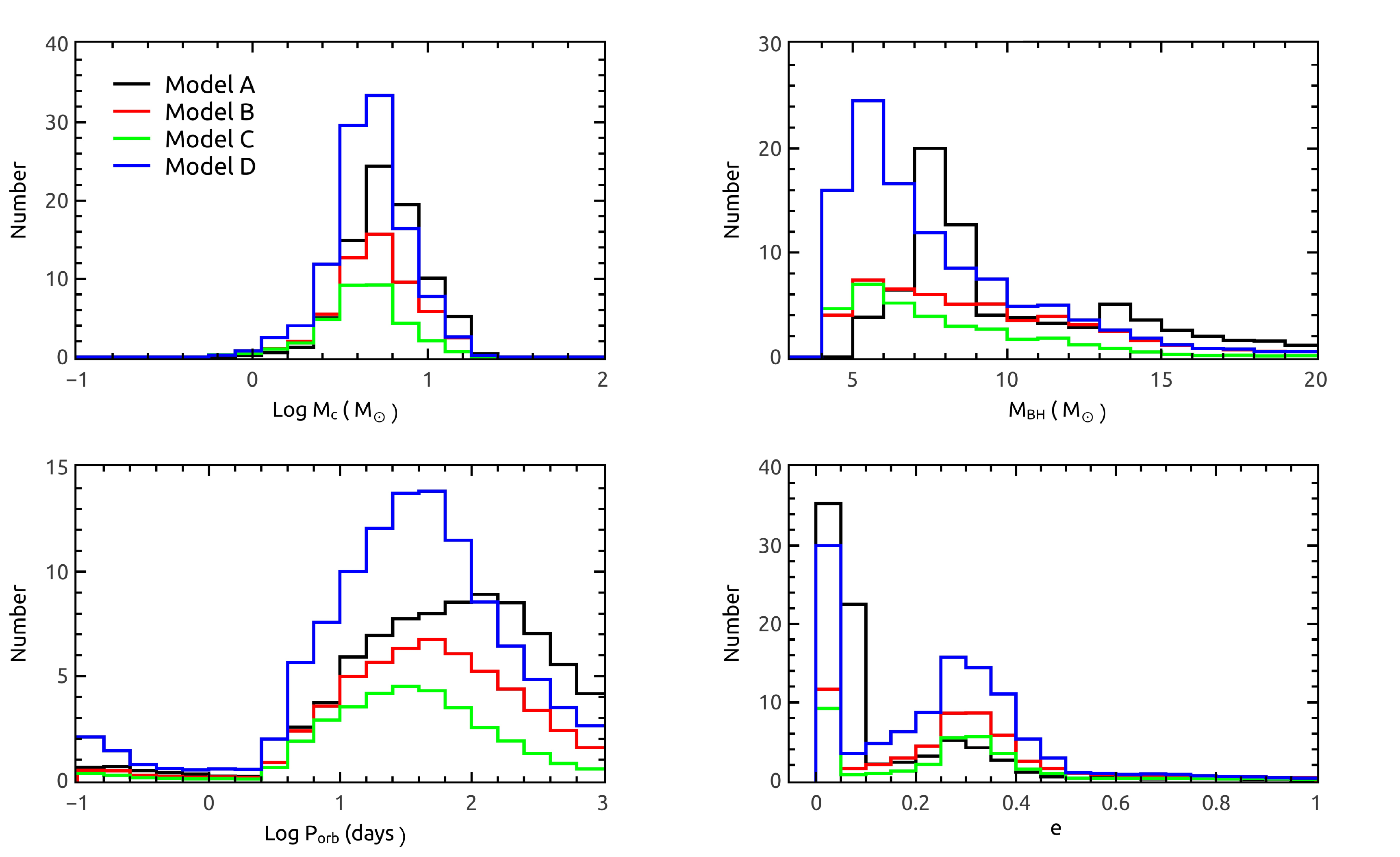}
\caption{Calculated number distributions of Galactic wind-fed BH$-$He XRBs 
as a function of the companion
mass, the BH mass, the orbital period, and the orbital eccentricity. 
   \label{figure6}}
\end{figure*}

\subsection{The Case of BH$-$OB XRBs}

Fig.~3 shows the predicted number distributions of Galactic BH$-$OB XRBs in 
the companion mass vs. orbital period plane for Models A$ - $D. Also the histogram diagrams of calculated number distributions as
a function of the binary parameters are plotted in Fig.~4. It can be seen that all our assumed models 
are characterized by similar distributions of the companion masses and
the orbital periods, but significantly different distributions of the BH masses and the orbital eccentricities.
Since a powerful stellar wind is required to activate an XRB, the peak of the companion mass distribution
of the wind-fed BH$-$OB XRBs is shifted to higher masses of $ \sim 8-13M_\odot $, compared to that of the BH$-$Be binaries.
The orbital period distribution of such XRBs has a peak at $ \sim 4  $ days in Model A and $ \sim 10 $ days in Models
B$ - $D, since Model A tends to produce more circular binaries than other models. 
Similar to the case of BH$-$Be systems, the BH mass distribution of the BH$-$OB XRBs has a peak near $ \sim 7-8M_\odot $ 
in Model A while more binaries with light ($ \sim 5M_\odot $ ) BHs are produced in Models B$ - $D. 
The source Cyg X-1 can be covered by the low probability region in the  
companion mass vs. orbital period plane, which may indicate that the majority of the BH$ - $OB XRBs
are relatively dim X-ray sources. 
Our models predict that there are $ \sim 10-30 $ wind-fed BH$ - $OB XRBs in the Milky Way.

\subsection{The Case of BH$-$He XRBs}

Fig.~5 presents the calculated number distributions of Galactic wind-fed XRBs with a helium star in 
the companion mass vs. orbital period plane. The four panels correspond to Models A$-$D.
In all models, the BH$-$He XRBs are obviously separated into two regions by an orbital period $\sim2-3  $ days. 
The reason is that  BH$-$He XRBs can be formed via two different channels, depending on
whether the progenitor binaries experienced a stable mass-transfer stage or a CE phase.   
The square symbol in each panel marks the location of the source Cgy X-3. 
This XRB possess the typical parameters of the post-CE systems with close orbits. 

In Fig.~6 we plot the histogram distributions of the calculated number of Galactic BH$-$He XRBs as a function of 
the companion mass, the BH mass, the orbital period, and the orbital eccentricity for Models A$-$D.
Since the progenitor systems containing a BH and an OB star possess similar distributions
of the OB star masses and the orbital periods in all models,  the BH$-$He XRBs also have similar 
distributions of the helium masses and the orbital periods under assumption of different models. 
The masses of the helium stars mainly distribute in the range of $ \sim 2-16M_\odot $, 
and the orbital periods distribute in a wide range ($ \sim 0.1-1000 $ days) with a peak at $ \sim 30-100  $ days. 
Some recent investigations showed that the mass-transfer process in the lobe-filling BH binaries 
is more stable than
previously expected \citep[e.g.,][]{p17,sl18}, the maximal mass ratio of the donor star to the BH for 
stable mass transfer can reach as high as $ \sim 6 $. Hence only a few BH$ - $He XRBs with orbital periods 
of $ \lesssim 1 $ day are produced in our calculations. Remarkably, such wind-fed XRBs 
in close orbits may be bright enough to be detected. In nearby galaxies, there also exist 
a couple of such close XRBs \citep[][and references therein]{ei15}. During the evolution of the progenitor
binaries, mass accretion of the BH can increase its mass. 
We see that the BH masses in the BH$-$He XRBs distribute in a broad range of $ \sim 5-20M_\odot $. 
Model A predicts that the BHs are more likely to possess the masses of $ \sim 7-8M_\odot $ since more binaries 
survive from the BH formation via direct collapse without natal kicks,
while Models B$ - $D tend to produce light BHs with mass distributions having a peak at $ \sim5 M_\odot $
due to the IMF.  Differently from the BH systems with an OB star,  almost all BH$ - $He XRBs have relatively 
low eccentricities of $ \lesssim 0.4$, whose distribution has two distinct peaks at $ \lesssim 0.1 $ and $ \sim 0.3 $. 
Tides and mass transfer between binary components tend to circularize the orbits during the
progenitor system evolution. Our obtained low eccentricities can coincide with the observations of Galactic WR$-$O
binaries \citep{vdh01} that are also post-mass transfer systems, although we do not
include a detailed treatment for the orbital evolution of mass-transferring eccentric binaries 
\citep[e.g.,][]{sw09,dk16}.

\section{Populations of RLO XRBs}

Only a few BH HMXBs have been observed as the wind-fed systems in the Milky Way, most of BH XRBs 
are actually the RLO LMXBs (see Table 1). The observational parameters of the BH LMXBs 
include the BH mass $ M_{\rm BH} $, the orbital period $ P_{\rm orb} $, 
the companion mass $ M_{\rm c}$, the surface effective temperature $ T_{\rm eff} $ of the companion stars,
the BH's spin parameter $ a_{\ast} $, and the mean mass-transfer rate $\dot{M}_{\rm tr} $.

\subsection{The Binary Evolution Calculations}

To explore the properties of the RLO XRBs, we use a combination of BPS and
detailed binary evolution calculations. The former can provide the birthrate distribution of 
the incipient BH binaries\footnote{It is possible that the BH systems with a helium star can appear as
RLO XRBs, but they should be very rare since the lifetime is very short \citep[$ \lesssim 10^{5} $ yr,][]{tlp15}. 
So we do not
consider the RLO XRBs with a helium star in our calculations.}. 
With the latter we can follow the subsequent evolution and especially track their evolutionary 
sequences during the RLO processes. We take the initial chemical composition of the companion stars 
to be $ X = 0.7 $, $ Y = 0.28 $, $ Z = 0.02 $. The BH is treated as a point mass and assumed to be 
initially non-rotating. Considering the mass range $(\sim 5-15M_\odot)  $ of known BHs in the 
Milky Way \citep{cj14},  for simplicity we only take two initial masses of 7 and $ 11M_\odot $ for BHs.
The incipient BH binaries obtained from the BPS calculations
are employed to guide the sets of the \textit{MESA} grid of initial binary parameters. 
We separately deal with the initial binaries with different BH masses, according to whether or not
the BHs in the incipient binaries are massive than $ 9 M_\odot $.
Under each BH mass, we evolve thousands of binary systems with initially different donor masses and 
orbital periods. All initial binaries are assumed to have circular orbits. 
There is a caveat that the incipient BH binaries actually have eccentric orbits due to mass losses and possible 
kicks at BH formation. We simply assume that the incipient BH binaries are quickly circularized by 
tidal interactions with the orbital angular momentum conserved, although this assumption is not always valid, 
especially for the systems with long orbital periods.
We compute the evolutionary tracks of incipient BH binaries in a grid of companion masses  
distributed over a range of $ \sim0.7-60M_\odot $ in logarithmic steps of 0.025 and 
orbital periods over a range of  $ \sim0.25-1000 $ days in logarithmic steps of 0.1.  
The birthrate of each specific binary 
can be obtained by weight of the ones of the incipient BH binaries residing in the corresponding grid interval. 

For a small fraction of the binary systems in this library,
the code breaks down when the evolution calculation just begins, since the initial orbital period 
is so short that the companion star has already filled its RL. 
For some binaries with very large mass ratio or evolved donor stars, the 
mass-transfer rates can rapidly rise up to $ \geq 0.1 M_{\odot} \,\rm yr^{-1} $, these
systems are expected to quickly enter a CE phase \citep[][]{p17,sl18}. In this case, we use the mass transfer rate of 
$ 0.1M_{\odot} \,\rm yr^{-1} $ as a criterion to decide whether the code should be terminated. 
Note that the systems with a higher mass-transfer rate can hardly contribute to the population of the RLO XRBs.

\begin{figure*}[hbtp]
\centering
\includegraphics[width=0.8\textwidth]{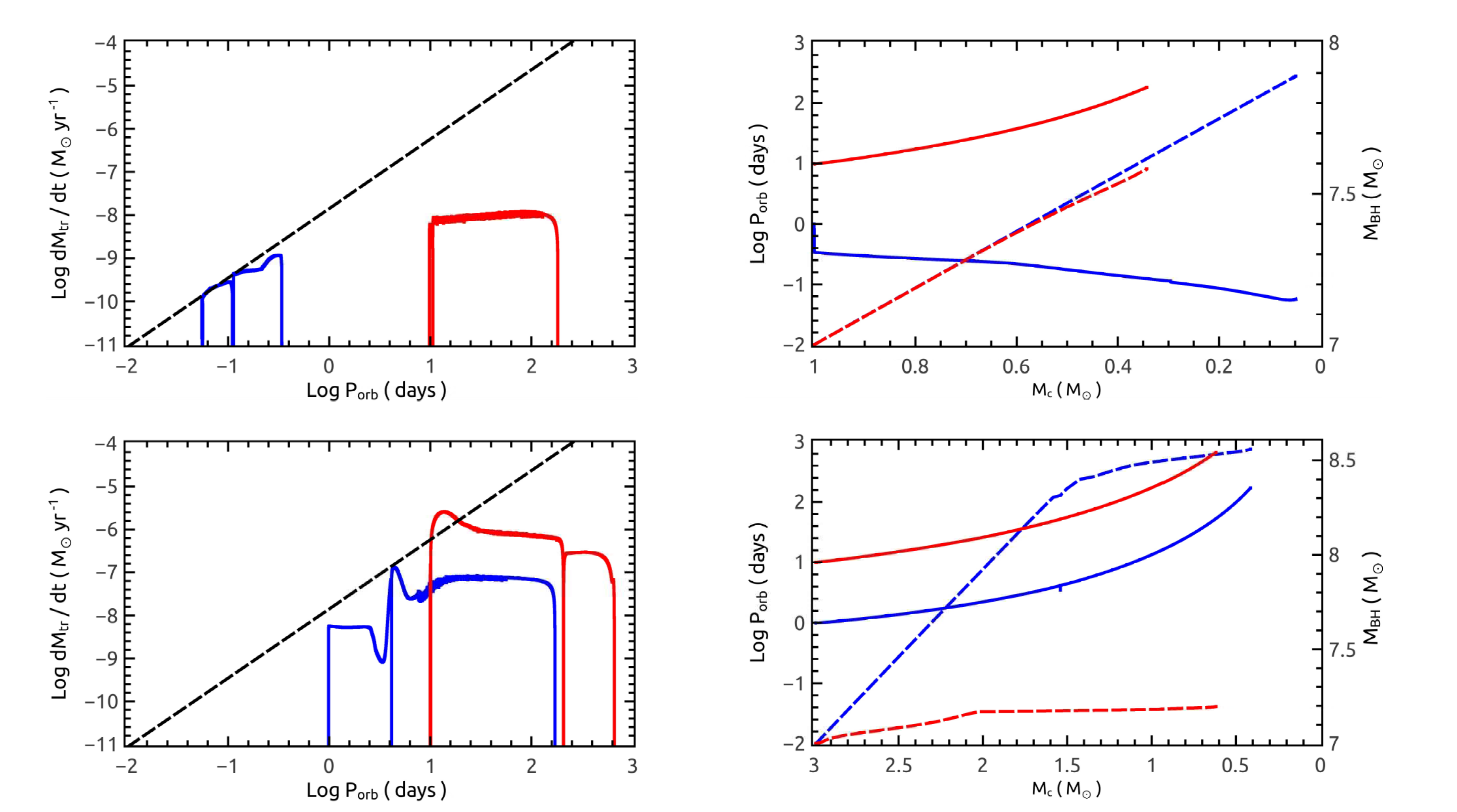}
\caption{Example evolution of the binary systems containing an initially  $7M_\odot $ BH. 
The top and bottom panels correspond to the initial companion masses of 1 and $ 3M_\odot $, 
the blue and red curves correspond to the initial orbital periods of 1 and 10 days, respectively. 
The evolutionary tracks of the mass-transfer rates as a function of the orbital periods 
are presented in the left panels, and the
orbital periods (solid curves) and the BH masses (dashed curves) as a function of the donor masses
in the right panels. The black
dashed lines reflect the critical mass-transfer rates for accretion disk instability.
   \label{figure7}}
\end{figure*}

The orbital evolution of a binary is driven by mass exchange and angular momentum losses 
due to gravitational wave radiation, magnetic braking, and possible matter ejection from the 
binary system. We use the formula of \citet{rvj83} for
magnetic braking, with the index $ \gamma $ set to be 3. This mechanism 
is assumed to work only for low-mass $ (\leq1.5M_\odot) $ stars with a convective envelope. 
We adopt the scheme of \citet{r88} to compute the mass-transfer rate 
$ \dot{M}_{\rm tr} $ via RLO. The mass accretion rate onto a BH is limited by the Eddington
accretion rate 
\begin{equation}
\dot{M}_{\rm E} = \frac{4\pi G M_{\rm BH}}{\eta\kappa c},
\end{equation}
 where $ G $ is the 
gravitational constant, $ \kappa $ the opacity, $ c $  the speed of light in vacuum, 
and $ \eta $ the efficiency of the BH in converting rest mass into radiative energy. This 
efficiency is approximately given by 
\begin{equation}
\eta = 1-\sqrt{1- \left( \frac{M_{\rm BH}}{3M_{\rm BH}^0 } \right)^{2}} 
\end{equation}
for $ M_{\rm BH} < \sqrt{6} M_{\rm BH}^0$, where  $ M_{\rm BH}^0 $ is the initial mass 
of the BH \citep{b70}. As the BH accretes 
mass and angular momentum, its spin parameter $ a_{\ast} $ is calculated according to 
\begin{equation}
a_{\ast} = \left(  \frac{2}{3}\right)^{1/2} \frac{M_{\rm BH}^0}{M_{\rm BH} } \left\lbrace 4-
\left[ 18\left(\frac{M_{\rm BH}^0}{M_{\rm BH} }\right) ^{2}-2\right] ^{1/2} \right\rbrace 
\end{equation}
for $ M_{\rm BH} < \sqrt{6} M_{\rm BH}^0$ \citep{t74}.

The matter flow from the companion star forms an accretion disk around the BH. 
Depending on whether or not the accretion disk is thermally and viscously unstable \citep{l01}, 
BH XRBs can be divided into transient and persistent sources. The disk state can be 
distinguished by a critical mass-transfer rate, for which we adopt the formula of \citet{ldk08} for irradiated accretion disk. 
If the mass-transfer rate falls below the critical value, the XRB is assumed to be a transient source 
experiencing short-lived outbursts separated by long-term quiescent phases. Matter accumulates in the disk during the 
quiescent phases and accretes during the outburst phases. The duty cycle is defined as the ratio of 
the outburst duration to the recurrence period, and we set it to be 0.03 \citep{k03} in our
calculations. For transient sources, the mass accretion rates can be greatly enhanced during outbursts. 
In this case, the accretion rate of 
the BH is also constrained by the Eddington limit.  Matter 
lost from the binary system is assumed to carry away the specific orbital angular momentum of the BH. 

\begin{figure*}[hbtp]
\centering
\includegraphics[width=0.75\textwidth]{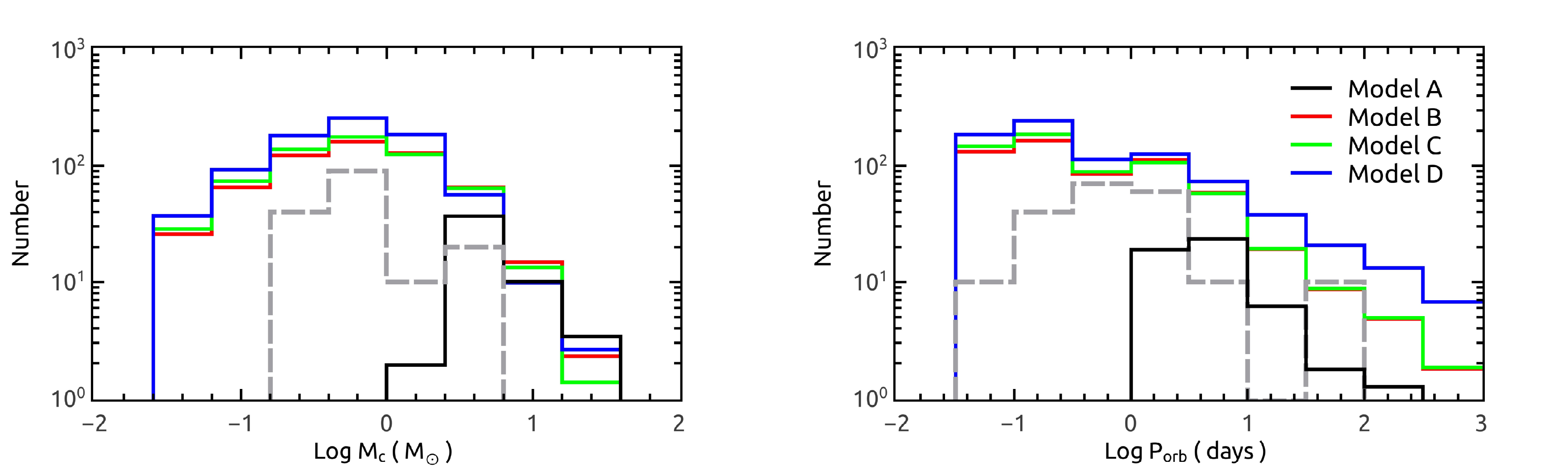}
\caption{Number distributions of Galactic RLO BH XRBs as a function of 
the companion mass (left panel) and the orbital period (right panel). 
The colored solid curves correspond to Models A$ - $D, and the gray dashed curves 
denote the observational data multiplied by a factor of 10. 
  \label{figure8}}

\end{figure*}

For sub-Eddington accretion, the X-ray luminosity is given as 
\begin{equation}
L_{\rm X}=\eta\dot{M}_{\rm tr}c^{2}.
\end{equation}
When $\dot{M}_{\rm tr} $ is larger than the Eddington 
rate $ \dot{M}_{\rm E} $,
the accretion disk may appear to be geometrically thick, which can significantly influence the X-ray luminosity. 
In this case, more and more of the material can be blown away as it flows inward to the BH,
so that the accretion disk never violates the Eddington limit locally \citep{ss73}.
In our calculations, we follow the method of \citet{kl16} to convert the mass-transfer rate into the X-ray luminosity. 
The total accretion luminosity, by integrating the local disk radiation \citep{ss73}, can be expressed as 
\begin{equation}
L_{\rm acc} \simeq L_{\rm E}\left[1+\ln \left(
\frac{\dot{M}_{\rm tr}}{\dot{M}_{\rm E}} \right) \right],
\end{equation}
where $ L_{\rm E} $ is the Eddington luminosity. According to this equation, the binary system can radiate an 
X-ray luminosity limited by several times the Eddington limit.
As a result of the geometric collimation, one can see the X-ray source in
directions within one of the radiation cones, so an apparent (isotropic) X-ray luminosity is
\begin{equation}
L_{\rm X} \simeq \frac{L_{\rm E}}{b}\left[1+\ln \left(
\frac{\dot{M}_{\rm tr}}{\dot{M}_{\rm E}} \right) \right],
\end{equation}
where $ b $ is the beaming factor. For this factor,
\citet{k09} proposed an approximate formula 
\begin{equation}
b \simeq \frac{73}{\dot{m}^2}, 
\end{equation}
where $ \dot{m} = \dot{M}_{\rm tr} / \dot{M}_{\rm E}$. This formula is valid when $  \dot{m} $ is greater than $ \sim 8.5 $, otherwise  
the beaming effect does not work (i.e., $ b = 1$). Accordingly, we assume that the potential of 
detecting an XRB along the beam is decreased by a factor of $ b $ in our calculations.

\begin{figure*}[hbtp]
\centering
\includegraphics[width=0.79\textwidth]{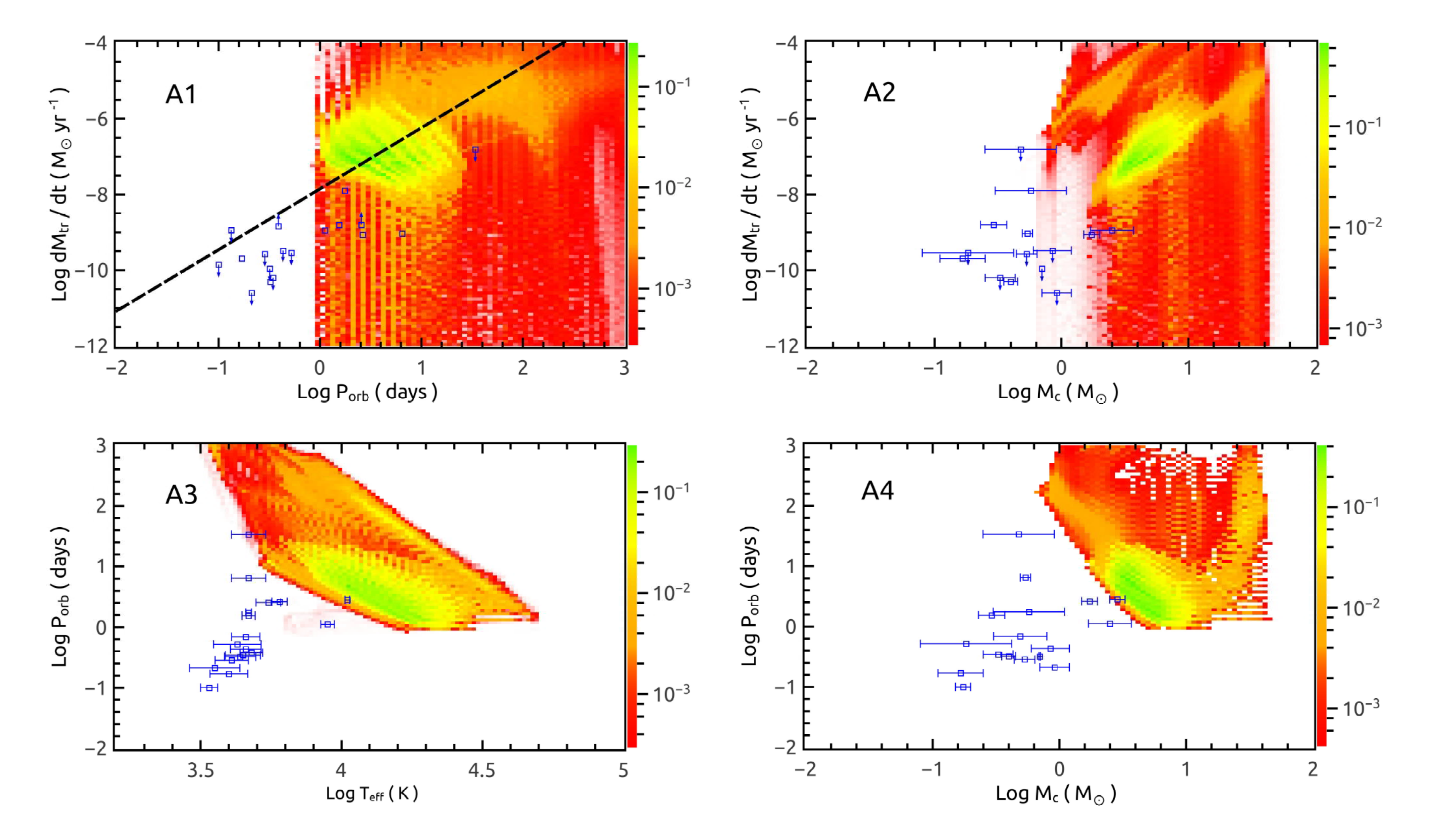}
\includegraphics[width=0.79\textwidth]{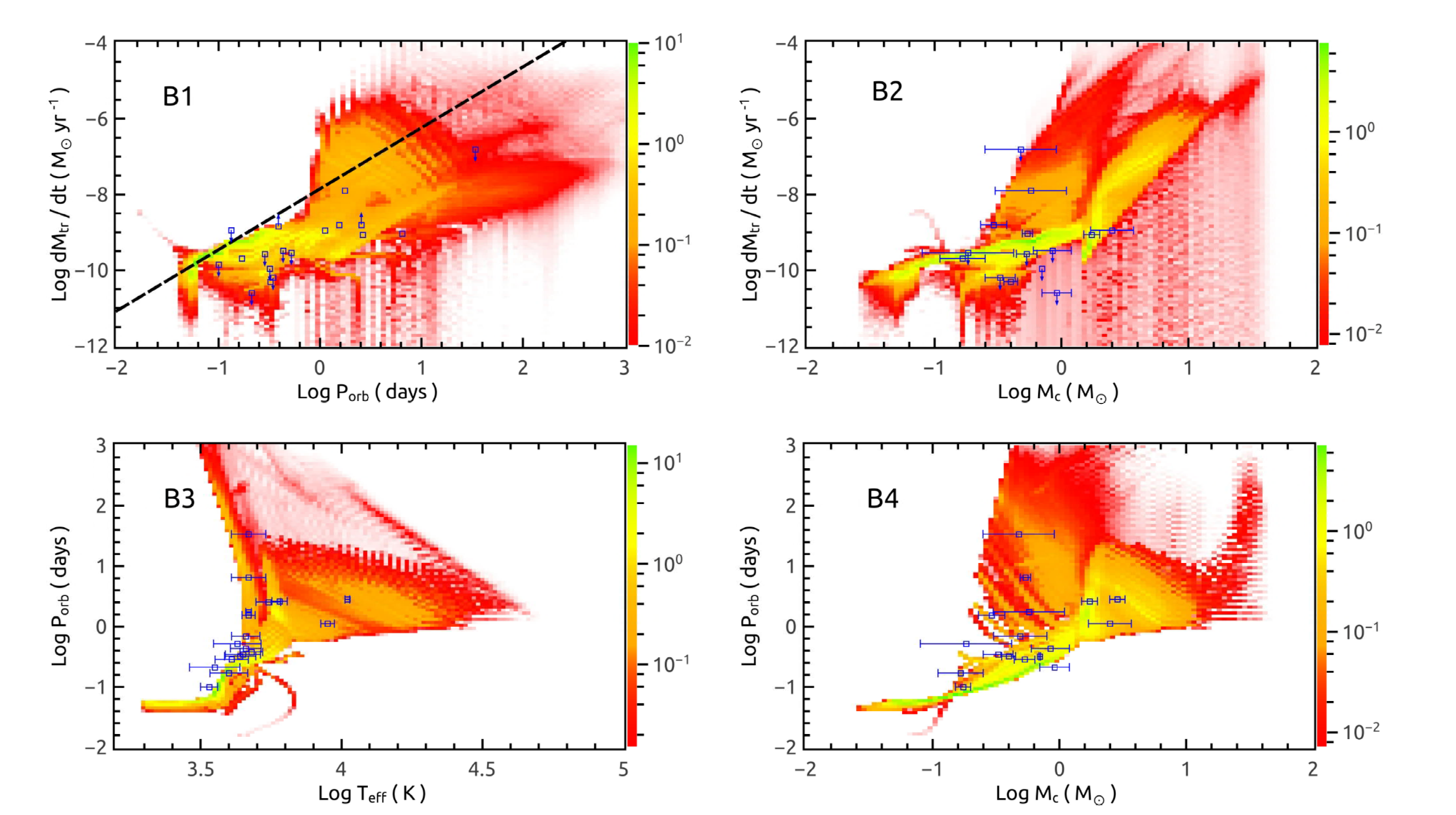}
\caption{Population diagrams for Galactic RLO BH XRBs in Model
A (top four panels) and B (bottom four panels). 
(1): Mass transfer rate vs. orbital period. (2): Mass transfer rate vs. companion mass. (3): Orbital period
vs. companion's surface effective temperature. (4): Orbital period vs. companion mass. 
The colors in each pixel are scaled according to the XRB numbers. 
The observational data are plotted with the blue squares.
The black dashed line in each panel (1) reflects the critical mass-transfer rates for accretion disk instability. 
   \label{figure9}}

\end{figure*}

\subsection{Evolution of a single binary}

Figure 7 shows sample evolutionary paths of four individual BH binaries, in which the initial mass of the BH is 
fixed to be $ 7M_\odot $. We take the initial masses of the 
companion stars to be 1 (top panels) and $ 3M_\odot $ (bottom panels), and the initial orbital periods 
to be 1 (blue curves) and 10 days (red curves). From this figure, there is a common tendency that the 
mass-transfer rates are higher for the systems with longer orbital
periods and larger companion masses. In the cases with a $ 3M_\odot $ companion star, 
as the material is transferred from the less massive companion to the more
massive BH,  the binary orbit enlarges. The companion star remains close to thermal equilibrium and
the mass transfer is driven by the nuclear evolution of the companion star. 
The binary finally evolves to be a wide LMXB. A similar evolutionary path can be found for the case with a 
$ 1M_\odot $  companion in a 10 day orbit. However, for
the binary initially containing a $ 1M_\odot $ companion in a 1 day orbit, the orbital evolution is mainly controlled by 
the mechanisms of magnetic braking and gravitational wave radiation, 
leading the binary to be a converging system. This is a typical scenario 
of forming a close BH LMXB. When the companion mass decreases to $ \sim 0.3M_\odot $ and the orbital period 
drops to $ \sim 0.1 $ day, the binary experiences a detached phase due to the turn-off of the magnetic braking 
for a fully convective star. We note that the mass-transfer rates in all binaries are usually below the critical rates 
for accretion disk instability, indicating that the BH LMXBs 
are likely transient X-ray sources. By considering the role of accretion disk instability and
the Eddington limit, a fraction of the transferred material is ejected out
of the binary system. This explains the difference in the final BH mass in different cases.

\begin{figure*}[hbtp]
\centering
\includegraphics[width=0.8\textwidth]{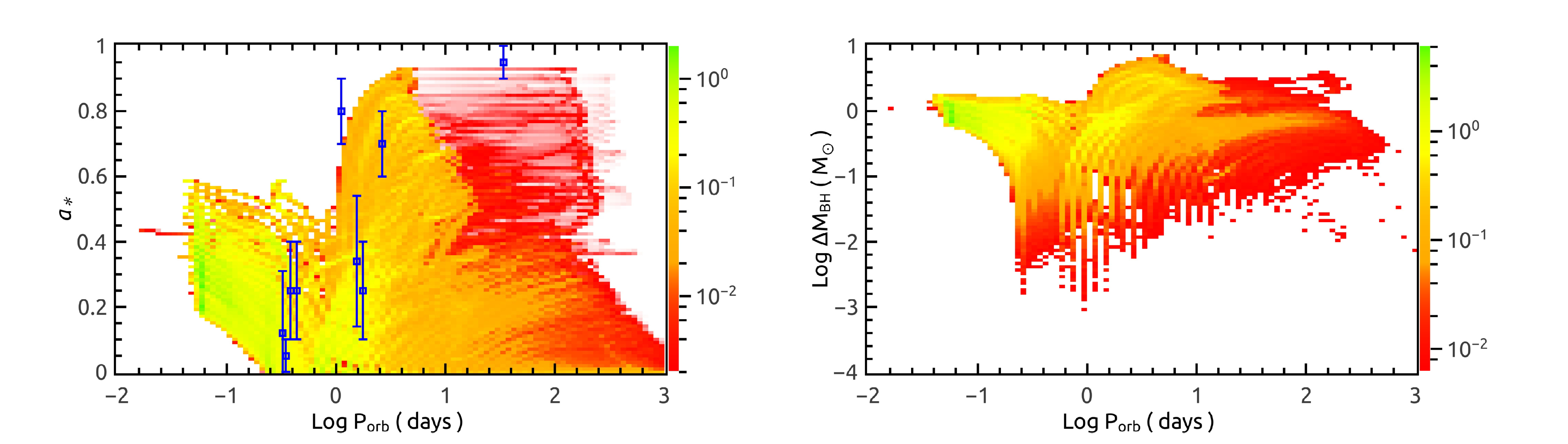}
\caption{Calculated number distributions of Galactic RLO XRBs in the $a_{\ast} - P_{\rm orb}  $
(left panel) and $\Delta M_{\rm BH}  - P_{\rm orb}  $ (right panel) planes under assumption of Model B. 
The colors in each pixel are scaled according to the XRB numbers. The blue squares mark 
the positions of observed BH LMXBs. 
   \label{figure10}}
\end{figure*}

\begin{figure*}[hbtp]
\centering
\includegraphics[width=0.8\textwidth]{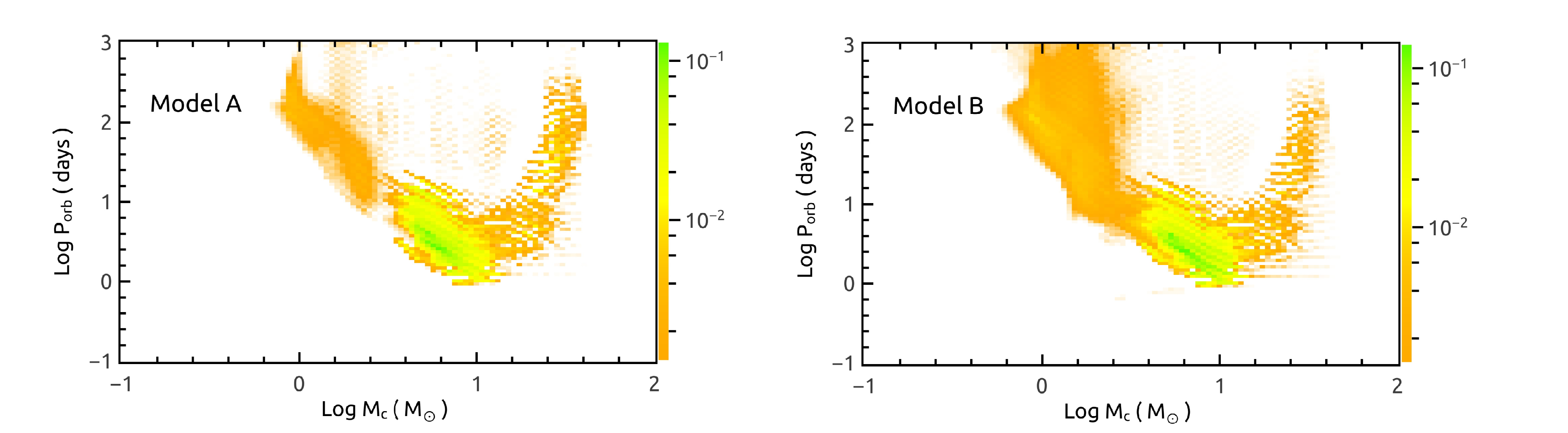}
\caption{Calculated number distributions of the BH ULX systems in the orbital period vs. companion mass plane. 
The left and right panels correspond to Model A and B, respectively. The colors are scaled according to the
ULX  numbers.
   \label{figure11}}

\end{figure*}

\subsection{Formation of BH LMXBs}

In Figure 8, we plot the calculated number distributions of the RLO XRBs with a BH accretor 
as a function of the companion mass (left panel)
and the orbital period (right panel) in Models A$-$D. The dashed curve in each panel denotes the distribution of 
the corresponding parameters of observed RLO XRBs in the Milky Way. 
As shown in Paper I, the incipient 
BH binaries with companion masses less than $ \sim 5M_\odot $ are hardly formed in Model A. Under the assumption
of this model, we obtain that the total number of the RLO XRBs is about 53 in the Milky Way. These XRBs have the 
donor masses of $ \gtrsim 1M_\odot $ and the orbital periods of $ \gtrsim 1 $ day, which cannot 
match the observations. In Models B$ - $D, the incipient BH binaries with a low-mass companion can be 
effectively created (see Paper I). Due to the mechanisms of magnetic braking and gravitational wave radiation, a fraction of them
can evolve to be the converging systems, leading to the formation of close BH LMXBs. 
Our calculations show that the overall 
population of the RLO XRBs has the number of several hundreds (see also Table 2), which is dominated by the LMXBs 
with donor masses distributing at a peak of
$\sim 0.6M_\odot $ and orbital periods distributing at a peak of $ \sim0.2 $ day. 
We can see that
the calculated distributions of the companion masses and the orbital periods roughly coincide 
with the observations of Galactic RLO XRBs in Models B$-$D. 
Since the donor masses and the orbital periods of the RLO XRBs have similar
distributions under assumption of Models B$-$D, we only show detailed parameter distributions of the XRB systems 
in Models A and B in the following.

Figure 9 shows the color images of the evolutionary tracks for all Galactic RLO XRBs
in Models A and B. Each panel contains a $ 100\times100 $ matrix element of the corresponding parameters. The colors
in each pixel are scaled according to the XRB numbers, which can be obtained by accumulating the 
product of the birthrates of the binaries passing through the corresponding matrix element with the time durations. 
The blue squares mark the positions of known RLO BH XRBs in the Milky Way. 
In panels A1 and B1, the dashed line reflects the critical mass-transfer rates for accretion disk instability \citep{ldk08}. 
One main difference between both models is that the minimal mass of BH progenitors drops to $ \sim 15 M_\odot $
in Model B compared to $\gtrsim 20 M_\odot $ in Model A. Stars below $ 20M_\odot $ have much lower stellar winds and tend to 
develop more classical giant structures at the end of their evolution, which is accompanied by much lower envelope 
binding energies. This mass decrease of BH progenitors can lead to not only
the reduction of the envelope masses, but also the dramatic decline of the envelope binding energies for CE evolution.
In Model B, the incipient BH binaries with a low-mass companion can survive the 
CE evolution, finally leading to the formation of LMXBs.
Since LMXBs cannot be formed in Model A, we focus on the calculated outcomes in Model B.
We can see that most of the RLO XRBs are expected to be
transient sources, since the mass-transfer rates are well below the critical values. 
The XRB systems mainly distribute in the 
region with mass transfer rates of $ \lesssim 10^{-9}M_\odot\,\rm yr^{-1} $ and orbital periods of $ \lesssim 1 $ day.
Panels B2 and  B4 show that the companion stars with masses $ \lesssim1M_\odot $ dominate the population 
of the RLO XRBs. This is because the low-mass companions in close binaries always stay on the main sequence,  
and experience long-lasting mass transfer phases driven by angular 
momentum loss due to magnetic braking and gravitational 
wave radiation. So the majority of the RLO XRBs are LMXBs. 
A small fraction of them evolve to be diverging systems with relatively long orbital periods, and
these binaries tend to have relatively high mass-transfer rates. From panel B3, we can see that
the effective temperatures of the companion star mainly distribute in the range of $ \sim 3000-9000\,\rm K $. 
Overall, the calculated distributions in Model B can roughly match the observations for the various parameters 
of the RLO XRBs. 

\begin{figure}[hbtp]
\centering
\includegraphics[width=0.5\textwidth]{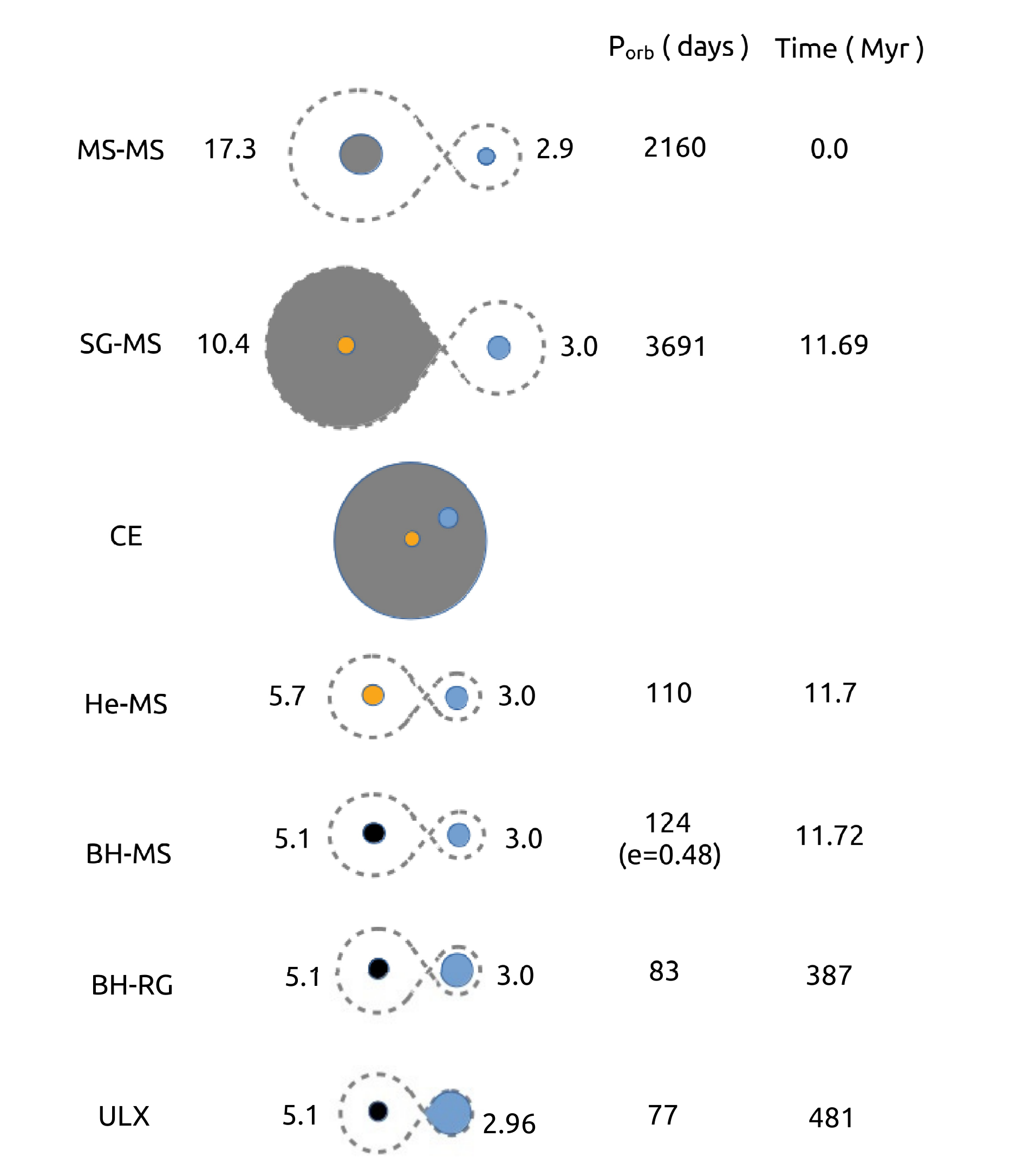}
\caption{A schematic plot depicts the formation of a BH binary similar to the one discovered by \citet{tk19}
and finally a ULX system. Acronyms for different stellar types used in this figure$ - $MS: main sequence; SG: supergiant; 
He: helium star; RG: red giant. 
   \label{figure12W}}
\end{figure}

It was proposed that the measured spins of BHs in LMXBs can be accounted for by the accreted matter,
even if the BHs are initially non-rotating \citep{prh03,fm15}. From the accreted mass
$ \Delta M_{\rm BH} $ we can obtain the spin parameters $ a_{\ast} $ for BHs in the RLO 
XRBs\footnote{The spins of BHs in wind-fed HMXBs are not involved in our calculations, 
the possible origin was recently investigated by \citet{qm19}.}. 
Fig.~10 presents calculated number distributions of Galactic RLO XRBs in the $a_{\ast} - P_{\rm orb}  $
(left panel) and $\Delta M_{\rm BH}  - P_{\rm orb}  $ (right panel) planes in Model B. 
The blue squares mark the positions of observed BH LMXBs. In each panel, there seem to be three high
probability regions corresponding to different types of evolutionary tracks. 
(1) For LMXBs with initial
$ P_{\rm orb} \lesssim 1$ day. 
$a_{\ast} $ gradually increases as the $P_{\rm orb}$ decreases, and $\Delta M_{\rm BH} $
can reach up to $ \sim 1.5 M_\odot $. Generally $a_{\ast}  \lesssim 0.6$. 
(2) For LMXBs with initial $ P_{\rm orb} \gtrsim 1$ day.  $a_{\ast} $  slowly increases
as the orbit enlarges, and $\Delta M_{\rm BH} $ is usually less than $ \sim 1 M_\odot $. 
Generally $ a_{\ast} \lesssim 0.4$.
(3) For IMXBs with initial $ P_{\rm orb} \sim  0.5-3 $ days. 
$a_{\ast} $ rapidly increases as the $P_{\rm orb}$ increases, and $\Delta M_{\rm BH} $
can reach as high as $ \sim 6 M_\odot $. In this case, $a_{\ast} $ can cover a wide range of $ \lesssim 0.9 $. 
Our calculations demonstrate that the spins of BHs in LMXBs can be explained by the transferred matter 
from the companion star, but the formation of GRS 1915+105 hosting a high spin BH is out of our expectation.
If GRS 1915+105 is the descendant of an IMXB, such a source is expected to be very short-lived and hardly detected. 
It is possible that GRS 1915+105 is evolved from an initially LMXB in which the BH was born with a high spin.

\begin{figure*}[hbtp]
\centering
\includegraphics[width=0.75\textwidth]{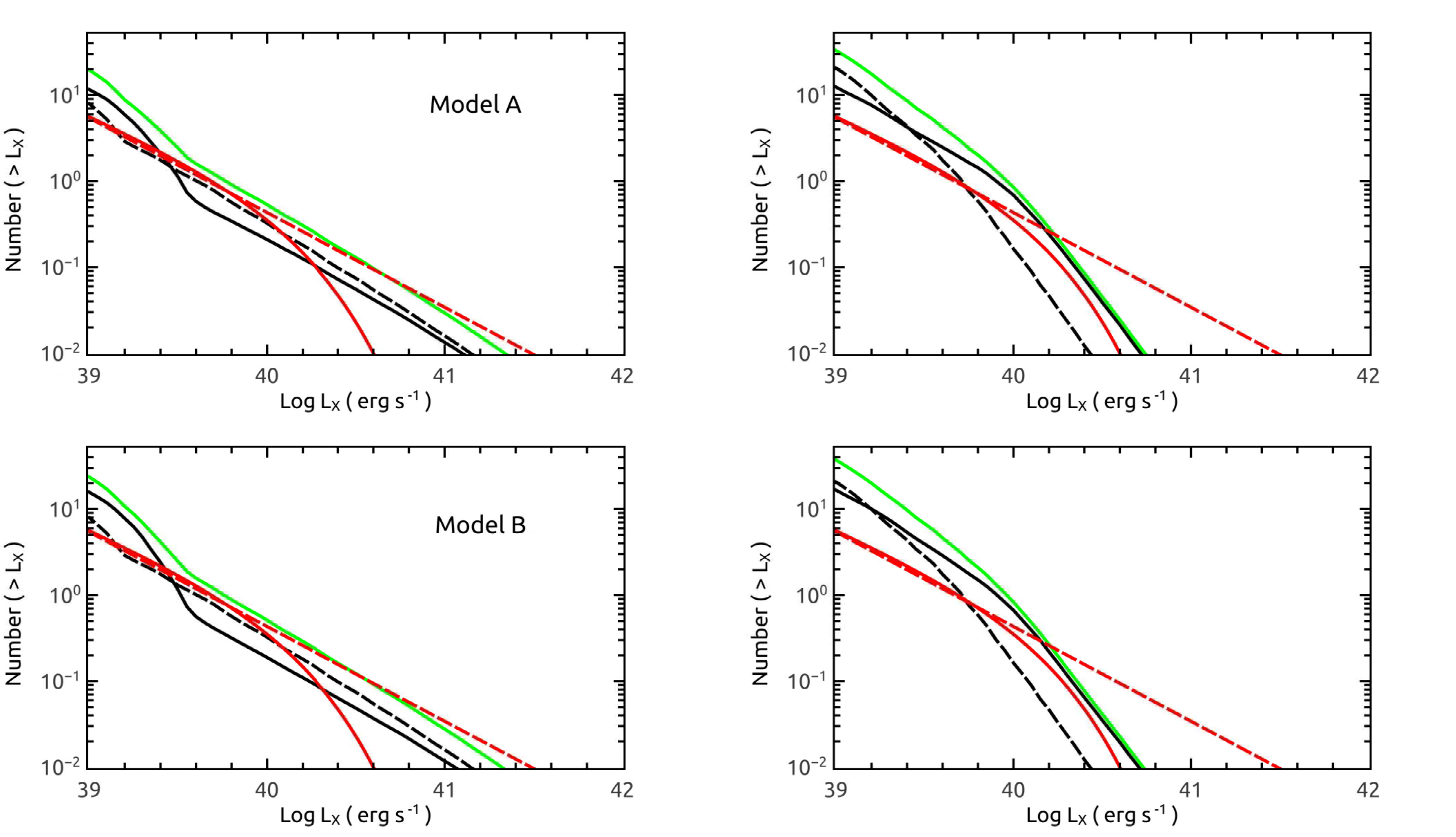}
\caption{The X-ray luminosity function for the ULX systems in a Milky Way$ - $like galaxy. The top and bottom
panels correspond to Model A and B, and the left and right panels correspond to the cases when using Equation~
(6) and (4) to calculate the X-ray luminosity, respectively. In each panel, the black solid and black dashed curves 
correspond to 
the ULX systems with a BH and an NS accretor, respectively. The green curve denotes the whole ULX population.
Based on the observed ULX sample in external galaxies,
the luminosity functions are fitted when applying two models 
of a power-law with an exponential cut-off (red solid curve) and a pure power-law (red dashed curve). 
Here the fit parameters are taken from \citet{s11}.
   \label{figure13}}

\end{figure*}

\subsection{Expected Population of BH ULXs}

Among the RLO XRBs, a small group of them can appear as ULXs (with X-ray luminosities greater than 
$ 10^{39} \rm\, erg\,s^{-1} $) when the mass-transfer rates are higher than 
$\sim 10^{-7}\, M_{\odot}\rm\,yr^{-1}$ \citep[e.g.,][]{kl16}. The ULX systems are expected to be so rare that they
are hardly detected in the Milky Way, but hundreds of them have been discovered in external galaxies due to their bright
X-ray emission. Figure 11 depicts the number distribution of the BH ULXs
in a Milky Way$ - $like galaxy in the companion mass vs. orbital period plane. The left and right panels
correspond to the calculated results in Model A and B, respectively. 
The effect of geometric beaming on the detection probability of a 
ULX has been considered. In Model A, the number of BH ULXs is $ \sim 12 $. Almost all 
of them contain an intermediate-mass donor, and the orbital periods 
distribute in the range of $ \sim1-10 $ days. Since the incipient BH LMXBs cannot 
be produced in Model A, the BH ULXs are mainly 
IMXBs. In Model B, the ULX number in a Milky Way$ - $like galaxy rises to $ \sim 16 $.
About 80\% ($ \sim13 $) of the ULX systems are IMXBs and the rest ($ \sim3 $) are LMXBs with $P_{\rm orb} \gtrsim 10 $ days,
in which the low-mass donors have climbed to the giant branches at an age older than 100 Myr. 
We can see that in both models there are few BH ULX systems with a massive 
donor of mass $ \gtrsim10M_\odot $.

Recently, \citet{tk19} discovered a non-interacting binary containing a $ \sim2.6-6.1 $ BH and 
a $ \sim 3 M_\odot$ giant star in a $ \sim 83 $ day orbit. The subsequent evolution of this 
binary must experience a ULX phase. Here we briefly discuss the formation and evolution of 
a binary system similar to this source \citep[see also][]{bca19}.  Since the incipient BH binaries with a
$ \sim 3 M_\odot$ companion cannot be produced in Model A, we only check the BPS outcomes
in Model B.  The companion mass and the orbital period of this source are used to pick out the 
required systems with similar binary parameters.  As an example (see Fig.~12), 
the evolution begins from a primordial binary consisting of a $ 17.3M_\odot $ primary and
a $ 2.9 M_\odot$ secondary in a 2160 day orbit. At the age of 11.7 Myr, the primary star that has evolved 
to be a supergiant star fills its RL and a CE phase follows. Shortly, the post-CE primary collapses into 
a BH with mass of $ 5.1 M_\odot $. At the moment, the remnant binary becomes a detached BH system 
containing a $ \sim 3M_\odot $ main-sequence star in an eccentric ($ e = 0.48 $) orbit with period of 124 days. At
the age of 387 Myr, the companion star has already climbed to the red giant branch, and the binary is 
circularized to have a circular orbit with period of 83 days. After the giant star fills its RL, this 
binary can appear as a ULX, whose age is about 481 Myr.  The discovery of such wide BH binary  
demonstrates that some BH ULXs with a low-mass donor are indeed existed in an old environment.

Recent observations have uncovered a population of ULXs with an NS accretor \citep[e.g.,][]{bh14,fw16,i17a,i17b}.
\citet{sld19} studied the parameter distribution and the number of the NS ULXs.
In the calculations presented here, we focus on the BH ULX systems.
The initial conditions and input parameters used in both studies are the same, so we
can directly compare the relative contribution of the ULXs with either an NS or a BH accretor.
Figure~13 shows the X-ray luminosity function of the ULX population in a Milky Way$ - $like galaxy. The top
and bottom panels correspond to Model A and B, respectively. In each panel, the black 
solid curve corresponds to the BH ULXs, the black dashed curve corresponds to the NS ULXs \citep[from][]{sld19},  
and the green solid curve corresponds to the whole (totally) population of the ULX systems. For the observed 
ULX sample in external galaxies, \citet{s11} fitted the luminosity functions 
that are plotted with the red curves in this figure. We should note
that the luminosity functions from both calculations and observations have been normalized to 
a star formation rate of $ 3 M_{\odot}\,\rm yr^{-1} $ for Milky Way$-$like galaxies. 
In the left panels, the luminosity function for the BH ULXs 
has a distinct break at the position of $L_{\rm X} \sim 4\times 10^{39} \rm\, erg\,s^{-1} $, 
which corresponds to a point that the beaming effect starts working. For the NS ULXs, 
the corresponding luminosity function has a similar break at $L_{\rm X} \sim 2\times 10^{39} \rm \,erg\,s^{-1} $
\citep[see also][]{sld19}. In the right panels, we plot the calculated luminosity function
without taking into account the beaming effect for comparison, instead using Equation~(4) to calculate the 
X-ray luminosity. It is indicated that BH ULXs are predominantly non-beamed XRBs \citep[see also][]{wl19}.
We can see that the calculated ULXs with
$L_{\rm X} \lesssim 4\times 10^{39} \rm\, erg\,s^{-1} $  are significantly more than the observed,
although the luminosity function at $L_{\rm X} \gtrsim 4\times 10^{39} \rm\, erg\,s^{-1} $ can roughly match the observations.
We estimate that the BH binaries can contribute a dozen ULXs in a Milky Way$ - $like galaxy 
(see also in Table 2), which have a considerable contribution to the whole ULX population compared to the NS binaries. 
Our calculations show that
the total number of the ULX population is about $ 20-30 $, while the 
observations indicate that only a few ULXs
have been already detected. One possible origin of this discrepancy is the selection effect against 
observing transient ULXs.

\begin{figure*}[hbtp]
\centering
\includegraphics[width=0.75\textwidth]{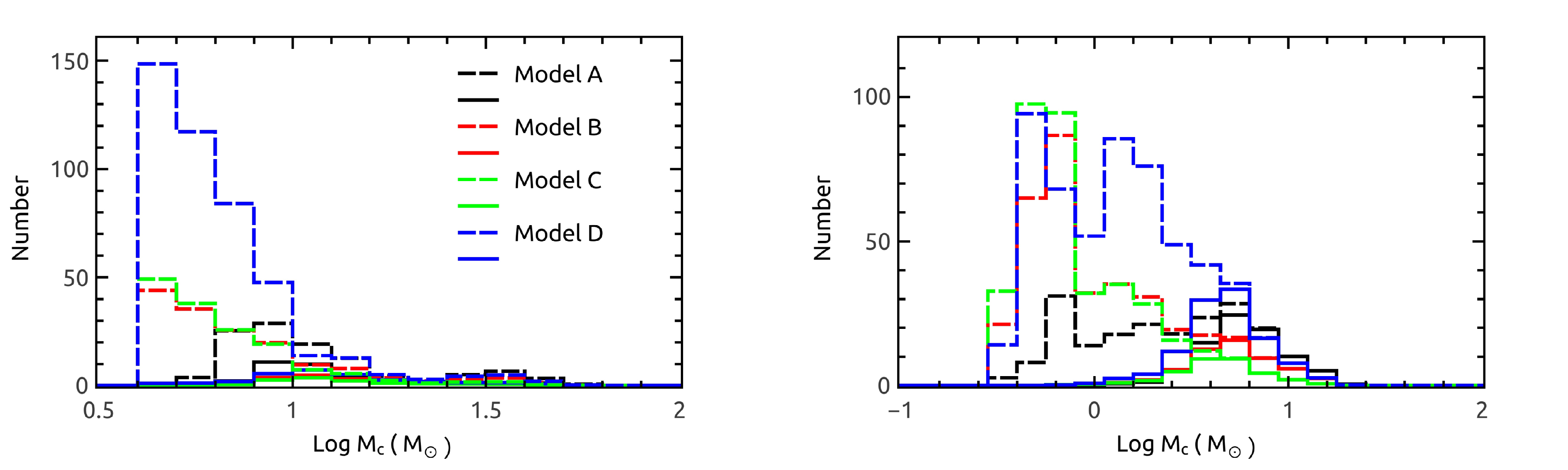}
\caption{Calculated number distributions of Galactic BH$-$OB (left panel) and BH$-$He (right panel) 
systems as a function of the companion masses in Models A$-$D. 
The dashed curves denote all detached 
systems, while the solid curves correspond to the wind-fed XRBs (same as those in Section 2). 
 \label{figure14}}
\end{figure*}

\section{Discussion}

For the wind-fed systems, we set an X-ray luminosity of $ 10^{35}\,\rm erg\,s^{-1} $ to pick out
the potential XRBs from our BPS calculations. Thus we obtain that the BH$ - $OB and BH$ - $He XRBs respectively
have the total numbers of $ \sim 10-30$ and $ \sim 30-110$ (see Table~2).  In fact, the
wind-fed XRBs with lower luminosities may be also detectable. Considering an extreme case that includes 
all detached BH binaries in spite of the X-ray luminosities (the BHs may have very low accretion rates),
we can provide an upper limit for the number of the wind-fed XRBs. Fig.~14 shows calculated number 
distributions of all detached BH$ - $OB (left panel) and BH$ - $He (right panel) binaries as a function of the
companion masses in Models A$-$D, which are plotted in the dashed curves. The number distributions
of Galactic wind-fed XRBs are also given for comparison, which are plotted in the
solid curves (same as those in Section 2). Under assumption of Models B$-$D, the mass spectrum of the 
normal-star companions in detached BH systems has a continuous distribution between 
$ \sim 0.7M_\odot $ and $ \sim 50M_\odot $ (see Paper I), here we set a cut-off mass of $ 4M_\odot $
to identify the OB stars (the detached systems with $ M_{\rm c} \leq 4M_\odot$ are the post-CE
systems evolved from the primordial binaries). For the BH$ - $OB binaries, we estimate that there are 
$ \sim 110-440 $ such detached systems in the Milky Way, most of them contain relatively 
less-massive B-type stars. 
For the BH$-$He binaries, our obtained total numbers in all models
are of the order 100, and the majority of them have a low-mass 
helium star. Evolved from the BH binaries with a giant companion, the binaries are identified as 
detached BH$ - $He systems
when the giant star losses its hydrogen envelope and becomes a low-mass helium star. 
This may happen just before the formation of a white dwarf, so the mass of the helium star can 
reach as low as $ \sim 0.4M_\odot $. Besides a normal-star companion (see Paper I), we emphasize that a small 
fraction of Galactic BH detached systems actually contain a helium star, which may be detected due to the 
optical emissions of the helium star.

\begin{figure*}[hbtp]
\centering
\includegraphics[width=0.75\textwidth]{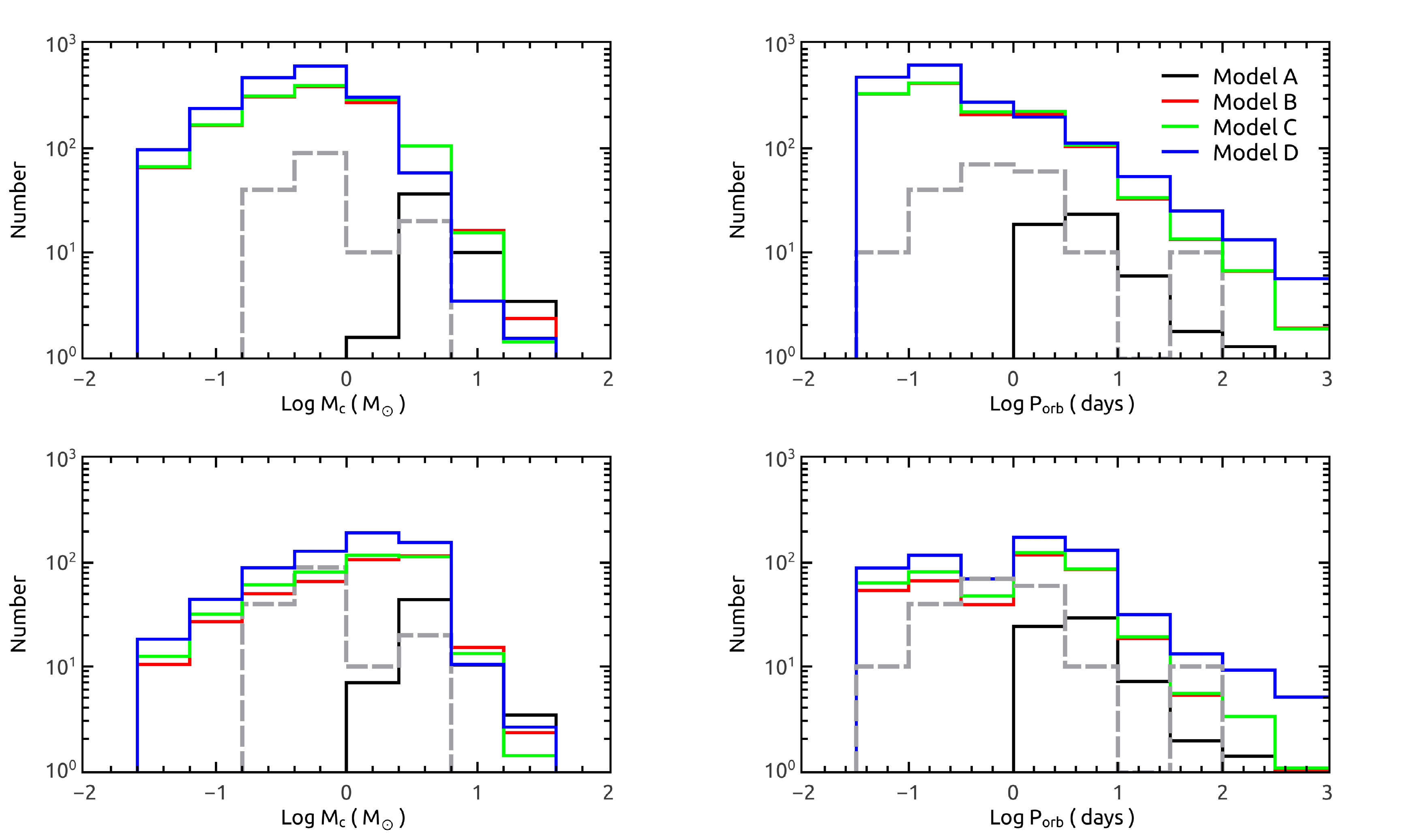}
\caption{Similar to Fig.~8 but taking the CE ejection efficiency to be
0.2 (top panels) and 5.0 (bottom panels). 
 \label{figure15}}
\end{figure*}

Next we briefly discuss the potential of detecting BH$ - $Be XRBs. 
So far, tens of BeXRBs have been confirmed to host an NS \citep{lvv06,r11}. Only MWC~656 and AS 386
are suggested to contain a BH orbited by a Be star via the optical observations of the Be star, none of them 
has been observed in the X-ray band. 
BPS studies suggest that BH$ - $Be binaries are much less than NS$ - $Be binaries \citep{bz09,sl14}. 
However, the estimated number ($ \sim 330-1040 $) of the BH$ - $Be binaries is about three times 
that ($ \sim 110-440 $) of the BH$ - $OB binaries (see Table 2). 
The source Cyg X-1 is an XRB with an O-type star, but no one BH$ - $Be XRB has be detected.
Recently, \citet{lss20} used a large grid of detailed binary evolution calculations to analyse 
the properties of the BH$ - $OB binaries in the Large 
Magellanic Cloud. It was mentioned by them that the majority of the BH$ - $OB binaries may be X-ray 
silent, even if quite a fraction of the systems actually contain a Be star companion.  Theoretical models usually
assume that a Be star is just a rapid rotator, but it is unclear that whether the Be disk can exist. 
One possibility
is that the Be disk has been disrupted when it closes to a WR star with a powerful wind 
during the evolution of the progenitor systems \citep{lss20}. 
Besides,  the truncation mechanism of the Be disk \citep{al94} may be so efficient that the accretion
rate of the BH is very low and the binary remains to be X-ray quiet \citep{zlw04}.

At last, we revisit the formation and evolution of BH LMXBs.  Evolved from massive primordial
binaries, the formation of LMXBs involves a CE phase. During the CE evolution, the orbital energy of the
embedded binary is assumed to eject the envelope. The magnitude of the CE ejection
efficiency has a big uncertainty, which can influence the birthrate and the parameter distribution of the incipient 
BH binaries. Note that we have set this efficiency to be unity in our previous calculations. When dealing with the post-CE binaries
with a white dwarf and a main-sequence star, \citet{zs10} indicated that the CE ejection efficiency 
should be in the range of $ 0.2-0.3 $. However, \citet{fa19} suggested a very high efficiency of $ \approx 5 $
when simulating the inspiral of an NS inside the envelope of a red supergiant\footnote{A CE ejection efficiency 
higher than unity obviously violates energy conservation during the CE evolution. However, for a
supergiant star losing mass rapidly during a CE phase, the reaction of the stellar radius to mass loss changes suddenly when 
a large fraction of the envelope has been stripped and the hydrogen abundance drops below a critical limit. At this point, 
the envelope of the supergiant contracts quickly and the binary system becomes detached. After the CE phase, some
hydrogen-rich materials still remain around the stellar core. This remaining hydrogen layer can dominate the overall
gravitational binding energy of the envelope due to its small radius, leading to an equivalently high value of the
CE ejection efficiency \citep{fa19}.}.  
In order to match the observed correlation between X-ray luminosity and displacement for the HMXB population,
\citet{zl14} inferred that a modest efficiency $ \sim 0.8-1.0 $ is more preferable. In Fig.~15 we show the calculated
number distributions of Galactic BH RLO XRBs as a function of the companion mass and the orbital period 
in Models A$-$D, when assuming the CE ejection efficiency to be
0.2 (top panels) and 5.0 (bottom panels). Obviously, the calculated distributions in Model A still cannot 
match the observations, since the BH systems with a low-mass companion cannot survive the CE evolution. 
In the top panels,  Models B$ - $D predict that the companion mass distribution has a peak near 
$0.6 M_\odot $ and the orbital period distribution has a peak near 0.2 day. The total number of the RLO XRBs in the Milky
Way is $ \sim 1300-1800$. In the bottom panels, the peaks of the companion mass and orbital period distributions
are respectively shifted to $ \sim 1M_\odot $ and  $ \sim 1  $ day, compared to the case with a smaller CE ejection efficiency. 
The calculated number of Galactic RLO XRBs in Models B$ - $D drops to $ \sim 390-650 $. 
This is because a large CE ejection efficiency tends to produce wide incipient BH binaries against the formation of the 
converging LMXBs. 
The evolution of BH LMXBs in our study is driven by the standard mechanisms of 
magnetic braking, gravitational wave radiation and possible mass ejection. 
Long-term observations show that some BH LMXBs are experiencing
extremely rapid orbital decays, which are greatly larger than expected if only including the conventional mechanisms
\citep{gh12,gh14,gh17}. Some exotic mechanisms are required to strongly extract the orbital angular momentum of the 
BH LMXBs, e.g., the interaction of the binary system with a surrounding circumbinary disk \citep{cl15,xl18,cp19}. 
The real mechanisms of driving the evolution of BH LMXBs may be very complicated, which are not further considered
in our simulations. Predictably, these uncertainties can significantly change the predicted properties and 
the number size of the BH LMXBs in the Milky Way.

\section{Conclusions}

Our previous study (Paper I) have simulated the population of Galactic detached BH binaries
with normal-star companions for a range of physical models. 
Model A involves the traditional mechanism for BH formation, in which the initial masses of BH progenitors 
are $ \gtrsim 20-25 M_\odot $. In Models B$-$D we assume that the BH progenitor masses can drop as low as
$ \sim 15 M_\odot $ while the newborn BHs possess different kick velocity distributions. In this work,
we further explore the potential population of BH XRBs, involving the number sizes and the parameter 
distributions. By use of 
both binary population synthesis and detailed binary evolution calculations, we  
obtain the predicted properties for various types of BH XRBs that include the wind-fed
systems and the RLO ones. The wind-fed XRBs are usually I/HMXBs containing an OB star
or a helium star around the BH. Our calculated outcomes of wind-fed XRBs do not strongly 
depend on the assumed models, in all of which the observed sources can be well reproduced. 
Model A predicts that the wind-fed XRBs are more likely to harbour a $ \sim 7-8M_\odot $ BH in
a nearly circular orbit, while Models B$-$D tend to produce the XRB systems with a light ($ \sim 5M_\odot $) BH 
in an eccentric orbit. The RLO XRBs are predominately L/IMXBs. 
By comparing our predictions with the observed properties of known BH LMXBs 
in the Milky Way, 
we find that the formation of LMXBs favors the models (i.e., Models B$-$D) involving 
relatively small masses for BH progenitors. 
Accordingly, we summarize our main results in Models B$ - $D as follows.

1. We estimate that there are totally $ \sim 500 - 1500 $ detached BH binaries with an OB star in the Milky Way, 
of which $\sim 3/4 $ are BH$ - $Be systems and $ \sim 1/4 $ are BH$ - $OB systems (see Table 2). 
Among the BH$ - $OB binaries, a small percent of them are expected to be 
detectable XRBs.  The source Cyg X-1
is such an XRB with an O-type star. MWC 656 and AS 386 are the BH$ - $Be binaries that were discovered by 
the optical observations of the Be star, but both of them are 
undetected in the X-ray band. In contrast to the NS$ - $Be XRBs, the majority of the BH$ - $Be systems may be
X-ray quiet due to the disruption of the Be disk \citep[see also][]{lss20}.

2. Our calculations show that the total number of Galactic wind-fed XRBs with a helium star is $ \sim30-110 $.
These XRBs can be generated via two different ways, depending on whether the progenitor systems 
experienced a stable mass-transfer stage or a CE phase. 
The post-CE systems are close XRBs, the candidate source Cyg X-3 is such a binary. 
The majority of the BH$ - $He XRBs are expected to be relatively wide systems, whose orbital period distribution has
a peak at $ \sim 30-100 $ days. 

3. Since the observability of the RLO XRBs depends on the mass-transfer timescale, the LMXBs with companion masses
of $ \lesssim 1M_\odot $ and orbital periods of $ \lesssim 1 $ day dominate the XRB population. 
The calculated distributions of various parameters of the RLO XRBs can roughly coincide with that of the observed
sample. We estimate that the number of the BH LMXBs in the Milky Way is  about several hundreds, provided that the 
mass transfer via RLO is driven by the standard mechanisms of the orbital angular momentum losses. When varying
the CE ejection efficiency in the range of $ 0.2-5.0 $, the total number of Galactic RLO XRBs can change by a factor of 
$ \sim 3 $. We find that the calculated LMXBs can better match the observations when adopting a relatively low 
efficiency of $ \lesssim 1 $.

4. The RLO XRBs may appear as ULXs if the mass transfer proceeds rapidly
at a high rate of $\gtrsim 10^{-7}\, M_{\odot}\rm\,yr^{-1}$. We estimate that the RLO systems with a BH accretor can 
contribute about a dozen ULXs in a Milky Way$ - $like galaxy, which have a comparable contribution with the NS binaries 
to the whole ULX population. The BH ULXs may exist in both young and old environments, which are respectively dominated 
by IMXBs and LMXBs.

\acknowledgements
We thank the referee for useful suggestions that helped
improve this paper.
This work was supported by the Natural Science Foundation 
of China (Nos. 11973026 and 11773015), the Project U1838201 
supported by NSFC and CAS, the National Program on Key Research and 
Development Project (Grant No. 2016YFA0400803), and 
the Fundamental Research Funds for the Central Universities (No. 14380033). 


\clearpage


\begin{thebibliography}{28}
\expandafter\ifx\csname natexlab\endcsname\relax\defNatureexlab#1{#1}\fi




\bibitem[{{Artymowicz \& Lubow}(1994)}]{al94}
Artymowicz, P., \& Lubow, S. H. 1994, ApJ, 421, 651

\bibitem[{{Bachetti} et~al.(2014)}]{bh14}
Bachetti, M., Harrison, F. A., Walton, D. J., et al. 2014, \nat, 514, 202

\bibitem[{{Bardeen}(1970)}]{b70}
Bardeen, J. M. 1970, Nature, 226, 64


\bibitem[{{Belczynski} et~al.(2008)}]{bk08}
Belczynski, K., Kalogera, V., Rasio, F., et al. 2008, ApJS, 174, 223

\bibitem[{{Belczynski \& Ziolkowski}(2009)}]{bz09}
Belczynski, K., \& Ziolkowski, J. 2009, ApJ, 707, 870

\bibitem[{{Belczynski} et~al.(2016)}]{bh16}
Belczynski, K., Holz, D. E., Bulik, T., \& O'Shaughnessy, R., 2016, Nature, 534, 512



\bibitem[{{Breivik} et~al.(2019)}]{bca19}
Breivik, K., Chatterjee, S., \& Andrews, J. J. 2019, ApJ, 878, L4
     



\bibitem[{{Brown}(2001)}]{b01}
Brown, G. E. 2001, NewAR, 6, 457



\bibitem[{{Calvelo} et~al.(2009)}]{cv09}
Calvelo, D. E., Vrtilek, S. D., Steeghs, et al. 2009, MNRAS, 399, 539

\bibitem[{{Cantrell} et~al.(2010)}]{cb10}
Cantrell, A. G., Bailyn, C. D., Orosz, J. A., et al. 2010, ApJ, 710, 1127

\bibitem[{{Casares \& Charles}(1994)}]{cc94}
Casares, J. \& Charles, P. A. 1994, MNRAS, 271, L5 

\bibitem[{{Casares} et~al.(2009)}]{coz09}
Casares, J., Orosz, J. A., Zurita, C., et al. 2009, ApJS, 181, 238

\bibitem[{{Casares} et~al.(2014)}]{cnr14}
Casares, J., Negueruela, I., Rib\'o, M., et al. 2014, Nature, 505, 378


\bibitem[{{Casares \& Jonker}(2014)}]{cj14}
Casares, J. \& Jonker, P. G. 2014, Spa. Sci. Rev., 183, 223

\bibitem[{{Chen \& Li}(2006)}]{cl06}
Chen, W.-C., \& Li, X.-D. 2006, MNRAS, 373, 305

\bibitem[{{Chen \& Li}(2015)}]{cl15}
Chen, W.-C., \& Li, X.-D. 2015, A\&A, 583, 108

\bibitem[{{Chen \& Podsiadlowski}(2019)}]{cp19}
Chen, W.-C., \& Podsiadlowski, P. 2019, ApJ, 876. L11


\bibitem[{{Chevalier \& Ilovaisky}(1990)}]{ci90}
Chevalier, C., \& Ilovaisky, S. A. 1990,  A\&A, 238, 163


\bibitem[{{Coriat} et~al.(2012)}]{cfd12}
Coriat, M., Fender, R. P., \& Dubus, G. 2012, MNRAS, 424, 1991

\bibitem[{{Corral-Santana} et~al.(2011)}]{cs11}
Corral-Santana, J. M., Casares, J., Shahbaz, T., et al. 2011, MNRAS, 413,L15









\bibitem[{{Dosopoulou \& Kalogera}(2016)}]{dk16}
Dosopoulou, F., \& Kalogera, V. 2016, ApJ, 825, 71

\bibitem[{{Eggleton \& Verbunt}(1986)}]{ev86}
Eggleton, P.P., \& Verbunt, F., 1986, MNRAS, 220, 13


\bibitem[{{Esposito} et~al.(2015)}]{ei15}
Esposito, P., Israel, G. L., Milisavljevic, D. et al. 2015, MNRAS, 452, 1112

\bibitem[{{Ertl} et~al.(2016)}]{ej16}
Ertl, T., Janka, H.-T., Woosley, S. E., Sukhbold, T., \& Ugliano, M. 2016, ApJ, 818, 124

\bibitem[{{Fabrika}(2004)}]{f04}
Fabrika, S. 2004, Astrophysics and Space Physics Reviews, 12, 1


\bibitem[{{Filippenko} et~al.(1999)}]{fl99}
Filippenko, A. V., Leonard, D. C., Matheson, T., et al. 1999, PASP, 111, 969

\bibitem[{{Fragos \& McClintock}(2015)}]{fm15}
Fragos, T., \& McClintock, J.E. 2015, ApJ, 800, 17

\bibitem[{{Fragos} et~al.(2019)}]{fa19}
Fragos, T., Andrews, J., Ramirez-Ruiz, E., et al. 2019, ApJL, 883, L45

\bibitem[{{Fryer} et~al.(2012)}]{f12}
Fryer, C., Belczynski, K., Wiktorowicz, G., et al. 2012, ApJ, 749, 91

\bibitem[{{F\"urst} et~al.(2016)}]{fw16}
F\"urst, F., Walton, D. J., Harrison, F. A., et al. 2016, ApJL, 831, L14

\bibitem[{{Gelino} et~al.(2001)}]{gh01}
Gelino, D. M., Harrison, T. E., \& McNamara, B. J. 2001, AJ, 122, 971

\bibitem[{{Giesers} et~al.(2018)}]{gd18}
Giesers, B., Dreizler, S., Husser, T.-O., et al. 2018, MNRAS, 475, L15



\bibitem[{{Gonz\'alez Hern\'andez} et~al.(2008)}]{gh08}
Gonz\'alez Hern\'andez,, J. I., Rebolo, R., \& Israelian, G. 2008,  A\&A, 478, 203

\bibitem[{{Gonz\'alez Hern\'andez} et~al.(2012)}]{gh12}
Gonz\'alez Hern\'andez, J. I., Rebolo, R., \& Casares, J. 2012, ApJL, 744, L25

\bibitem[{{Gonz\'alez Hern\'andez} et~al.(2014)}]{gh14}
Gonz{\'a}lez Hern{\'a}ndez, J. I., Rebolo, R., \& Casares, J. 2014, MNRAS, 438, L21

\bibitem[{{Gonz\'alez Hern\'andez} et~al.(2017)}]{gh17}
Gonz\'alez Hern\'andez, J. I., Su{\'a}rez-Andr{\'e}s, L., Rebolo, R., \& Casares, J. 2017, MNRAS, 465, L15

\bibitem[{{Gou} et~al.(2010)}]{gm10}
Gou, L., McClintock, J. E., Steiner, J. F., Narayan, R., Cantrell, A. G.,
Bailyn, C. D., \& Orosz, J. A. 2010, ApJ, 718, L122

\bibitem[{{Gou} et~al.(2014)}]{gmr14}
Gou, L., McClintock, J. E., Remillard, R. A., et al. 2014, ApJ, 790, 29

\bibitem[{{Greiner} et~al.(2001)}]{gcm01}
Greiner, J., Cuby, J. G., \& McCaughrean, M. J. 2001, Nature, 414, 522



\bibitem[{{Hamann} et~al.(1995)}]{ham95}
Hamann, W.-R., Koesterke, L., \& Wessolowski, U., 1995, A\&A, 299, 151

\bibitem[{{Harlaftis} et~al.(1996)}]{hh96}
Harlaftis, E. T., Horne, K., \& Filippenko, A. V. 1996, PASP, 108, 762

\bibitem[{{Harlaftis} et~al.(1997)}]{hs97}
Harlaftis, E. T., Steeghs, D., Horne, K., \& Filippenko, A. V. 1997, AJ, 114, 1170

\bibitem[{{Harlaftis} et~al.(1999)}]{hc99}
Harlaftis, E., Collier, S., Horne, K., \& Filippenko, A. V. 1999, A\&A, 341, 491

\bibitem[{{Harlaftis \& Greiner}(2004)}]{hg04}
Harlaftis, E. T., \& Greiner, J. 2004,  A\&A, 414, L13



\bibitem[{{Hobbs} et~al.(2005)}]{h05}
Hobbs, G., Lorimer, D. R., Lyne, A. G., \& Kramer, M. 2005, MNRAS, 360, 974


\bibitem[{{Hurley} et~al.(2002)}]{h02}
Hurley, J. R., Tout, C. A., \& Pols, O. R. 2002, MNRAS, 329, 897


\bibitem[{{Hynes} et~al.(2003)}]{hsc03}
Hynes, R. I., Steeghs, D., Casares, J., et al. 2003, ApJ, 583, L95

\bibitem[{{Hynes} et~al.(2009)}]{hbr09}
Hynes, R. I., Bradley, C. K., Rupen, M., et al. 2009, MNRAS, 399, 2239

\bibitem[{{Ioannou} et~al.(2004)}]{ir04}
Ioannou, Z., Robinson, E. L., Welsh, W. F., \& Haswell, C. A. 2004, AJ, 127,481

\bibitem[{{Ivanova}(2006)}]{i06}
Ivanova, N., 2006, ApJ, 653, L137

\bibitem[{{Israel} et~al.(2017a)}]{i17a}
Israel, G. L., Belfiore, A., Stella, L., et al. 2017a, Science, 355, 817

\bibitem[{{Israel} et~al.(2017b)}]{i17b}
Israel, G. L., Papitto, A., Esposito, P., et al. 2017b, MNRAS, 466, L48


\bibitem[{{Johannsen} et~al.(2009)}]{jp09}
Johannsen, T., Psaltis, D., \& McClintock, J. E. 2009, ApJ, 691, 997

\bibitem[{{Justham} et~al.(2006)}]{j06}
Justham, S., Rappaport, S., Podsiadlowski, P., 2006, MNRAS, 366, 1415


\bibitem[{{Kaaret} et~al.(2017)}]{kfr17}
Kaaret, P., Feng, H., \& Roberts, T. P. 2017, ARA\&A, 55, 303

\bibitem[{{Kalogera}(1999)}]{k99}
Kalogera, V., 1999, ApJ, 521, 723


\bibitem[{{Khargharia} et al.(2010)}]{kfr10}
Khargharia, J., Froning, C. S., \& Robinson, E. L. 2010, ApJ, 716, 1105

\bibitem[{{Khargharia} et al.(2013)}]{kfr13}
Khargharia, J., Froning, C. S., Robinson, E. L., \& Gelino, D. M. 2013, ApJ, 145, 21

\bibitem[{{Khokhlov} et al.(2018)}]{km18}
Khokhlov, S. A., Miroshnichenko, A. S., Zharikov, S. V., et al. 2018, ApJ, 856, 158

\bibitem[{{Kiel \& Hurley}(2006)}]{kh06}
Kiel, P. D., \& Hurley, J. R., 2006, MNRAS, 369, 1152





\bibitem[{{King}(2002)}]{k02}
King, A. R. 2002, MNRAS, 335, L13

\bibitem[{{King} et al.(2003)}]{k03}
King, A. R., Rolfe, D. J., Kolb, U., \& Sshenker, K. 2003, MNRAS, 341, L35

\bibitem[{{King}(2009)}]{k09}
King A. R., 2009, MNRAS, 393, L41

\bibitem[{{King \& Lasota}(2016)}]{kl16}
King A. R., \& Lasota, J.-P. 2016, MNRAS, 458, L10




\bibitem[{{Kochanek}(2014)}]{k14}
Kochanek, C. S., 2014, ApJ, 785, 28


\bibitem[{{Kruckow} et~al.(2018)}]{kt18}
Kruckow, M. U., Tauris, T. M., Langer, N., Kramer, M., \& Izzard, R. G. 2018, MNRAS, 481, 1908

\bibitem[{{Kuulkers} et~al.(2013)}]{kk13}
Kuulkers, E., Kouveliotou, C., Belloni, T. et al. 2013, A\&A, 552, 32

\bibitem[{{Langer}(2012)}]{l12}
Langer, N. 2012, ARA\&A, 50, 107

\bibitem[{{Langer} et~al.(2020)}]{lss20}
Langer, N., Sch{\''u}rmann, C., Stoll, K. et al. 2020, A\&A, 638, 39

\bibitem[{{Lasota}(2001)}]{l01}
Lasota, J.-P. 2001, NewAR, 45, 449

\bibitem[{{Lasota} et~al.(2008)}]{ldk08}
Lasota, J.-P., Dubus, G., \& Kruk, K. 2008, A\&A, 486, 523

\bibitem[{{Lee} et~al.(1991)}]{los91}
Lee, U., Osaki, Y., \& Saio, H. 1991, MNRAS, 250, 432

\bibitem[{{Li}(2008)}]{l08}
Li, X.-D. 2008, MNRAS, 384, L16

\bibitem[{{Li}(2015)}]{l15}
Li, X.-D. 2015, NewAR, 64, 1

\bibitem[{{Linden} et~al.(2012)}]{lvk12}
Linden, T.; Valsecchi, F., \& Kalogera, V. 2012, ApJ, 748, 114

\bibitem[{{Liu} et~al.(2019)}]{liu19}
Liu, J., Zhang, H., Howard, A. et al. 2019, \nat, 575, 618

\bibitem[{{Liu} et~al.(2006)}]{lvv06}
Liu, Q. Z., van Paradijs, J., \& van den Heuvel, E. P. J. 2006, A\&A, 455, 1165

\bibitem[{{MacDonald} et~al.(2014)}]{mbb14}
MacDonald, R. K. D., Bailyn, C. D., Buxton, M., et al. 2014, ApJ, 784, 2

\bibitem[{{Macias} et~al.(2011)}]{mo11}
Macias, P., Orosz, J. A., Bailyn, C. D., et al. 2011, in Bulletin of the American Astronomical Society, Vol. 43, 
American Astronomical Society, 217, 143.04

\bibitem[{{Mapelli \& Giacobbo}(2018)}]{mg18}
Mapelli, M., \& Giacobbo, N. 2018, MNRAS, 479, 4391

\bibitem[{{Madhusudhan} et~al.(2008)}]{mr08}
Madhusudhan, N., Rappaport, S., Podsiadlowski, Ph., \& Nelson, L. 2008, ApJ, 688, 1235



\bibitem[{{McClintock} et~al.(2006)}]{msn06}
McClintock, J. E., Shafee, R., Narayan, R., et al. 2006, ApJ, 652, 518

\bibitem[{{McGill} et~al.(2013)}]{msj13}
McGill, M. A., Sigut, T. A. A., \& Jones, C. E. 2013, ApJS, 204, 2


\bibitem[{{Motta} et~al.(2014)}]{mbs14}
Motta, S. E., Belloni, T. M., Stella, L., et al. 2014, MNRAS, 437, 2554

\bibitem[{{Mu{\~n}oz-Darias} et~al.(2008)}]{mcm08}
Mu{\~n}oz-Darias, T., Casares, J., \& Mart{\'i}nez-Pais, I. G. 2008, MNRAS, 385, 2205

\bibitem[{{Neilsen} et~al.(2008)}]{ns08}
Neilsen, J., Steeghs, D., \& Vrtilek, S. D. 2008, MNRAS, 384, 849




\bibitem[{{O'Connor \& Ott}(2011)}]{oo11}
O'Connor, E., \& Ott, C. D. 2011, ApJ, 730, 70

\bibitem[{{Orosz} et~al.(1998)}]{oj98}
Orosz, J. A., Jain, R. K., Bailyn, C. D., McClintock, J. E., \& Remillard,R. A. 1998,  ApJ, 499, 375

\bibitem[{{Orosz} et~al.(2001)}]{okv01}
Orosz, J. A., Kuulkers, E., van der Klis, M., et al. 2001, ApJ, 555, 489

\bibitem[{{Orosz} et~al.(2009)}]{os09}
Orosz, J. A., Steeghs, D., McClintock, J. E., et al. 2009, ApJ, 697, 573

\bibitem[{{Orosz} et~al.(2011a)}]{om11}
Orosz, J. A., McClintock, J. E., Aufdenberg, J. P., et al. 2011a, ApJ, 742, 84

\bibitem[{{Orosz} et~al.(2011b)}]{os11}
Orosz, J. A., Steiner, J. F., McClintock, J. E., et al . 2011b, ApJ, 730, 75



\bibitem[{{Paczy{\'n}ski}(1976)}]{p76}
Paczy{\'n}ski, B. 1976, in Proc. IAU Symp. 73, Structure and Evolution in Close
Binary Systems, ed. P. P. Eggleton, S. Mitton \& J. Whealan (Dordrecht:
Reidel), 75



\bibitem[{{Pavlovskii} et~al.(2017)}]{p17}
Pavlovskii, K., Ivanova, N., Belczynski, K., \& Van, K. X., 2017, MNRAS, 465, 2092

\bibitem[{{Paxton} et~al.(2011)}]{p11}
Paxton, B., Bildsten, L., Dotter, A., et al. 2011, ApJS, 192, 3

\bibitem[{{Paxton} et~al.(2013)}]{p13}
Paxton, B., Cantiello, M., Arras, P., et al. 2013, ApJS, 208, 4

\bibitem[{{Paxton} et~al.(2015)}]{p15}
Paxton, B., Marchant, P., Schwab, J., et al. 2015, ApJS, 220, 15

\bibitem[{Pejcha \& Thompson(2015)}]{pt15}
Pejcha, O., \& Thompson, T. A. 2015, ApJ, 801, 90


\bibitem[{{Podsiadlowski} et~al.(1992)}]{pjh92}
Podsiadlowski, P., Joss, P. C., \& Hsu, J. J. L. 1992, ApJ, 391, 246

\bibitem[{{Podsiadlowski} et~al.(1995)}]{p95}
Podsiadlowski, P., Cannon, R.C., Rees, M.J., 1995, MNRAS, 274, 485


\bibitem[{{Podsiadlowski} et~al.(2003)}]{prh03}
Podsiadlowski P., Rappaport S., \& Han Z., 2003, MNRAS, 341, 385

\bibitem[{{Podsiadlowski} et~al.(2010)}]{p10}
Podsiadlowski, P., Ivanova, N., Justham, S., Rappaport, S., 2010, MNRAS, 406, 840



\bibitem[{{Portegies Zwart} et~al.(1997)}]{p97}
Portegies Zwart, S.F., Verbunt, F., \& Ergma, E., 1997, A\&A, 321, 207


\bibitem[{{Qin} et~al.(2019)}]{qm19}
Qin, Y., Marchant, P., Fragos, T., et al. 2019, ApJ, 870, L18

\bibitem[{{Raithel} et~al.(2018)}]{rs18}
Raithel, C. A., Sukhbold, T., \& \"Ozel, F. 2018, ApJ, 856, 35

\bibitem[{{Rappaport} et~al.(1983)}]{rvj83}
Rappaport, S., Verbunt, F., \& Joss, P. C. 1983, ApJ, 275, 713

\bibitem[{{Rappaport} et~al.(2005)}]{rpp05}
Rappaport, S., Podsiadlowski, Ph., \& Pfahl, E. 2005, MNRAS, 356, 401

\bibitem[{{Reig}(2011)}]{r11}
Reig, P. 2011, Ap\&SS, 332, 1

\bibitem[{{Remillard} et~al.(1996)}]{ro96}
Remillard, R. A., Orosz, J. A., McClintock, J. E., \& Bailyn, C. D. 1996, ApJ, 459, 226

\bibitem[{Remillard \& McClintock(2006)}]{rm06}
Remillard, R. A., \& McClintock, J. E. 2006, ARA\&A, 44, 49

\bibitem[{{Reynolds} et~al.(2007)}]{rc07}
Reynolds, M. T., Callanan, P. J., \& Filippenko, A. V. 2007, MNRAS, 374,657

\bibitem[{{Ritter}(1988)}]{r88}
Ritter, H 1988, A\&A, 202, 93





\bibitem[{{Sadakane} et~al.(2006)}]{saa06}
Sadakane, K., Arai, A., Aoki, W., et al. 2006, PASJ, 58, 595


\bibitem[{{Sepinsky} et~al.(2009)}]{sw09}
Sepinsky, J. F., Willems, B., Kalogera, V., \& Rasio, F. A. 2009, ApJ, 702, 1387

\bibitem[{{Smartt}(2009)}]{s09}
Smartt, S. J. 2009, ARA\&A, 47, 63

\bibitem[{{Smartt}(2015)}]{s15}
Smartt, S. J. 2015, PASA, 32, 16



\bibitem[{{Shafee} et~al.(2006)}]{smr06}
Shafee, R., McClintock, J. E., Narayan, R., et al. 2006, ApJ, 636, L113

\bibitem[{{Shahbaz} et~al.(1997)}]{sn97}
Shahbaz, T., Naylor, T., \& Charles, P. A. 1997, MNRAS, 285, 607

\bibitem[{{Shakura \& Syunyaev}(1973)}]{ss73}
Shakura, N. I., \& Syunyaev, R. A. 1973,  A\&A, 24, 337


\bibitem[{{Shao \& Li}(2014)}]{sl14}
Shao, Y., \& Li, X.-D. 2014, ApJ, 796, 37



\bibitem[{{Shao \& Li}(2018)}]{sl18}
Shao, Y., \& Li, X.-D. 2018, MNRAS, 477, L128

\bibitem[{{Shao \& Li}(2019)}]{sl19}
Shao, Y., \& Li, X.-D. 2019, ApJ, 885, 151

\bibitem[{{Shao} et~al.(2019)}]{sld19}
Shao, Y., Li, X.-D. \& Dai, Z.-G.  2019, ApJ, 886, 118



\bibitem[{{Steiner} et~al.(2011)}]{sr11}
Steiner, J. F., Reis, R. C., McClintock, J. E., et al. 2011, MNRAS, 416, 941

\bibitem[{{Steiner} et~al.(2011)}]{sm13}
Steiner, J. F., McClintock, J. E., \& Narayan, R. 2013, ApJ, 762, 104

\bibitem[{{Sukhbold} et~al.(2016)}]{se16}
Sukhbold, T., Ertl, T., Woosley, S. E., Brown, J. M., \& Janka, H.-T. 2016, ApJ, 821, 38



\bibitem[{{Swartz} et~al.(2011)}]{s11}
Swartz, D., Soria, R., Tennant, A., \& Yukita, M. 2011, ApJ, 741, 49




\bibitem[{{Tauris} et~al.(2015)}]{tlp15}
Tauris, T., Langer, N., \& Podsiadlowski, P. 2015, MNRAS, 451, 2123


\bibitem[{{Thompson} et~al.(2019)}]{tk19}
Thompson, T. A., Kochanek, C. S., Stanek, K. Z., et al. 2019, Science, 366, 637

\bibitem[{{Thorne}(1974)}]{t74}
Thorne, K. S. 1974, ApJ, 191, 507



\bibitem[{{Torres} et~al.(2019)}]{tc19}
Torres, M. A. P., Casares, J., Jim{\'e}nez-Ibarra, F. 2019, ApJ, 882, L21

\bibitem[{{Torres} et~al.(2020)}]{tc20}
Torres, M. A. P., Casares, J., Jim{\'e}nez-Ibarra, F. 2020, ApJ, 893, 37





\bibitem[{{Ugliano} et~al.(2012)}]{uj12}
Ugliano, M., Janka, H.-T., Marek, A., \& Arcones, A. 2012, ApJ, 757, 69



\bibitem[{{van den Heuvel}(1974)}]{vh74}
van den Heuvel, E. P. J. 1974, ApJ, 198, L109



\bibitem[{{van der Hucht}(2001)}]{vdh01}
van der Hucht, K. A. 2001, NewAR, 45, 135


\bibitem[{{van Kerkwijk} et~al.(1992)}]{vk92}
van Kerkwijk, M. H., Charles, P. A., Geballe, T. R., et al. 1992, Nature, 355, 703

\bibitem[{{Vinciguerra} et~al.(2020)}]{vnv20}
Vinciguerra, S., Neijssel, C. J., Vigna-G{\'o}mez, A. et al. 2020, arXiv: 2003.00195

\bibitem[{{Vink} et~al.(2001)}]{vink01}
Vink, J. S., de Koter, A., \& Lamers, H. J. G. L. M. 2001, A\&A, 369, 574



\bibitem[{{Wang} et~al.(2016)}]{wjl16}
Wang, C., Jia, K., \& Li, X.-D. 2016, MNRAS, 457, 1015



\bibitem[{{Webbink}(1984)}]{w84}
Webbink, R. F. 1984, ApJ, 277, 355





\bibitem[{{Wiktorowicz} et~al.(2017)}]{ws17}
Wiktorowicz, G., Sobolewska, M., Lasota, J.-P., \& Belczynski, K. 2017, ApJ, 846, 17

\bibitem[{{Wiktorowicz} et~al.(2019)}]{wl19}
Wiktorowicz, G., Lasota, J.-P., Middleton, M.,\& Belczynski, K. 2019, ApJ, 875, 53

\bibitem[{{Woosley \& Weaver}(1995)}]{ww95}
Woosley, S. E., \& Weaver, T. A., 1995, ApJS, 101, 181


\bibitem[{{Xu \& Li}(2010a)}]{xl10a}
Xu, X.-J., \& Li, X.-D. 2010a, ApJ, 716, 114

\bibitem[{{Xu \& Li}(2010b)}]{xl10b}
Xu, X.-J., \& Li, X.-D. 2010b, ApJ, 722, 1985


\bibitem[{{Xu \& Li}(2018)}]{xl18}
Xu, X.-T., \& Li, X.-D. 2018, ApJ, 859, 46



\bibitem[{{Yamaoka} et~al.(2012)}]{ya12}
Yamaoka, K., Allured, R., Kaaret, P., et al. 2012, PASJ, 64, 32



\bibitem[{{Zdziarski} et~al.(2013)}]{zmb13}
Zdziarski, A. A., Mikolajewska, J., Belczynski, K. 2013, MNRAS, 429, L104

\bibitem[{{Zhang} et~al.(2004)}]{zlw04}
Zhang, F., Li, X.-D., \& Wang, Z.-R. 2004, ApJ, 603, 663

\bibitem[{{Zorotovic} et~al.(2010)}]{zs10}
Zorotovic, M., Schreiber, M. R., G{\"a}nsicke, B. T., \& Nebot G{\'o}mez-Mor{\'a}n, A. 2010, A\&A, 520, 86

\bibitem[{{Zuo \& Li}(2014)}]{zl14}
Zuo, Z.-Y. \& Li, X.-D. 2014, MNRAS, 442, 1980

\bibitem[{{Zurita} et~al.(2002)}]{zs02}
Zurita, C., S{\'a}nchez-Fern{\'a}ndez, C., Casares, J., et al. 2002, MNRAS, 334, 999

\end{thebibliography}
\end{document}